\documentclass{aa}
\usepackage{graphicx}
\begin{document}
   \title{An inverse method to recover the star formation
    history and reddening properties of a galaxy from its spectrum.}
   \titlerunning{Inversion of galaxy spectra.}

   \subtitle{}

   \author{J.-L. Vergely 
          \inst{1}
          \and
          A. Lan\c{c}on
          \inst{1} 
          \and
          M. Mouhcine
          \inst{1}
          }

   \offprints{A.\,Lan\c{c}on}

   \institute{Observatoire Astronomique de Strasbourg (UMR 7550), 
   11 rue de l'Universit\'e, F--67000 Strasbourg, France
              }

   \date{Received ; accepted }

   \abstract{ 
   We develop a non-parametric inverse method 
   to investigate the star formation rate, the metallicity evolution and 
   the reddening properties of galaxies based on their spectral energy 
   distributions (SEDs).
   This approach allows us to clarify the level of information 
   present in the data, depending on its signal-to-noise ratio (S/N). 
   When low resolution SEDs are available in the ultraviolet, optical and 
   near-IR wavelength ranges together,
   we conclude that it is possible to constrain the star formation rate and 
   the effective dust optical depth simultaneously with a signal-to-noise 
   ratio of 25. With excellent signal-to-noise ratios, the age-metallicity
   relation can also be constrained.\\
   We apply this method to the well-known nuclear starburst in
   the interacting galaxy NGC\,7714. We focus on deriving the SFR and the 
   reddening law. We confirm that classical extinction models 
   cannot provide an acceptable simultaneous fit of the SED and the lines.
   We also confirm that, with the adopted population synthesis 
   models and in addition to the current starburst, 
   an episode of enhanced star formation that started  
   more than 200\,Myr ago is required. As the time elapsed since
   the last interaction with NGC\,7715, based on dynamical studies,
   is about 100\,Myr, our result reinforces the suggestion
   that this interaction might not have been the most important 
   event in the life of NGC\,7714.
   }
 
   \maketitle
%

\section{Introduction}
The integrated spectra of galaxies contain 
information about the ages of their stellar populations, the metallicity 
of the stars and the effects of dust extinction.

In order to infer the history of the star formation rate (SFR),
one usually tries to match the spectral energy distribution (SED) 
as closely as possible with model populations computed 
with various scenarios for the SFR. The quality of the SED fit is 
assessed either qualitatively by visual inspection or quantitatively, 
e.g. by the $\chi^2$-test or a more general expression of the likelihood.
Often, the SFR is represented by a number 
of discrete values or by a predefined analytic form. This approach, which could 
be called the direct or synthetic method, has been largely used in
the field of population synthesis (see Heavens et al. 2000, Reichardt et
al. 2001 for efficient recent implementations). 
Since one usually keeps the number of free parameters as small 
as possible, the adopted functional forms for the SFR and age-metallicity 
relation (AMR) impose certain limitations on what type of model 
populations can be considered. Typically, these will be combinations
of instantaneous bursts and episodes of constant star formation. 
Some inverse methods (Craig and Brown 1986, Tarantola and 
Valette 1982\,a,b) deal with such a problem in an opposite way : one 
tries to determine the functional form of e.g. the SFR with as much 
freedom as possible, with a resolution in time 
that is dictated by the information contained in the data.
Because of the latter property, these methods are called non-parametric.

This work presents a non-parametric inverse method to  estimate 
characteristics of galaxy evolution such as the SFR, the AMR and 
the intrinsic dust extinction. As for all such approaches, 
the method is based on a probabilistic formulation of inverse problems
(in our case the formalism of Tarantola \& Valette 1982\,a,b).
We apply the method in the framework of {\em evolutionary} population
synthesis. In other words, possible solutions are only sought among 
those compatible with our current understanding of
star formation and evolution: the relative fraction of stars
of various masses is not arbitrary but follows an initial
mass function, and theoretical evolutionary tracks combined with
stellar spectral libraries determine the possible emission spectra of 
isochrone stellar populations. 
A probabilistic formulation for the alternative
{\em empirical} population synthesis 
has been developed recently in the parametric case by Cid Fernandes et al. 
(2001). As noted by these authors, 
the exploration of solutions to an inversion problem can be 
tackled as a minimization problem or with an adequate sampling
algorithm for the space of parameters. The second type of approach
provides a complete description of the uncertainties
on the estimated parameters, but may become difficult to implement
in practice when some of the unknowns are non-parametric functions of time, 
which can take an immense variety of shapes.  Here a minimization procedure
is adopted. In addition, specific tools are used in order 
to estimate the validity of the inverse procedure, like the a posteriori 
covariance and resolution, and the mean index.

The nature of the available data determines many of the 
capabilities and limitations of inversion procedures. Sets of equivalent
widths in the optical spectrum have been used
because these measurements are relatively easy to acquire (Pelat 1997,
Boisson et al. 2000). Cid Fernandes et al.
(2001) combined equivalent widths and colours in order also to constrain
dust extinction. However, they used a classical one
parameter description of extinction and reddening. The effects of
dust are rarely that simple (Witt \& Gordon 2000). The 
work we present here was partly motivated by previous 
studies of starburst galaxy spectra, which not only made it clear
that average obscuration laws depend on the type 
of galaxy observed (Calzetti et al. 1994), but also showed that 
complex distributions of the stars and the dust in space
can lead to significant local deviations from this average. Lan\c{c}on
et al. (2001) studied the $\sim$330 central parsecs of the
nuclear starburst galaxy NGC\,7714, and emphasized the effects of
the different optical depths of dust along various lines of sight
within their small aperture. They demonstrated that in dusty
objects it is necessary to combine SEDs and emission lines
from the optical, the ultraviolet and the near-IR spectral ranges
if one aims at recovering the SFR over a wide range of ages and
useful information on the wavelength dependent attenuation by dust.
With those results in mind, we chose to apply the inversion method 
to data sets such as those of Lan\c{c}on et al. (2001): in this
paper, the empirical constraints are combined low resolution
SEDs in the three spectral ranges together with emission lines of H{\sc ii}. 
Computation times were not prohibitive in the present case. 

The paper is organised as follows. Section 2 \ref{mod} presents our model
assumptions for the SEDs and reddening. Section \ref{inv} 
introduces the inverse method.  After applying this technique to simulated 
SEDs in Sect.\,\ref{simul}, we give new 
constraints on the SFR and the reddening law for the nuclear
starburst of NGC 7714 in Sect.\,\ref{ngc}.\\  

\section{Modeling galaxy spectra.}
\label{mod}

\subsection{The basis $B_{\lambda}(t)$}

The intrinsic SED of a synthetic
stellar population depends on the following model ingredients :
\begin{itemize} 
   \item the SFR $\psi(t)$, which gives the total mass of stars 
   of current age $t$, 
   \item the initial mass function (IMF),
   \item the metallicity of the stars, $Z(t)$, which reflects the 
    metallicity of the interstellar medium from which the stars are born,
   \item the results of stellar evolution calculations,
   \item stellar atmosphere models, which give the flux at each wavelength 
   for any star,
   \item photoionization models, which add the nebular emission to the
   SED. 
\end{itemize}

Without extinction, the simulated flux distribution $F_{\lambda}$ 
is written as follows:
\begin{equation}
F_{\lambda}=\int_{t_f}^{t_i} \psi(t)\, B_{\lambda}(t,Z(t))\, {\rm d}t
\end{equation}
As $\psi(t)$ is always positive, it will be convenient to
write 
\begin{equation}
\psi(t) = \psi_o\,\exp(\alpha(t))
\label{alphadef.eq}
\end{equation} 
where $\psi_o$ is a constant (arbitrary, but fixed).
$B_{\lambda}(t,Z(t))$ is the spectrum of a coeval stellar population
of current age $t$. We considered ages between $t_f=1$\,Myr and 
$t_i=16$\,Gyr.
This model basis implicitly assumes a certain initial
mass function and rests on a particular set of stellar evolution
and stellar atmosphere models. In the following, we will test 
the IMFs of Scalo (1998) and of Salpeter (1955), 
with a lower stellar mass limit of 0.1\,M$_{\odot}$
and an upper mass limit of 120\,M$_{\odot}$. We construct basis spectra for a 
grid of ages and metallicities with the population synthesis code 
{\sc P\'egase} (Fioc \& Rocca-Volmerange, 1997). They are based on 
the stellar evolutionary tracks of the Stellar Astrophysics group of the 
Astronomical Observatory of Padua (and extensions thereof),
and on the semi-empirical spectral library of Lejeune et al. (1997, 1998).
We refer to Fioc \& Rocca-Volmerange (1997) for details.
Time is sampled with logarithmic timesteps of $\sim$0.1; 
the applications in the following
sections will show this is fine enough to not limit the resolution of 
the estimated evolutionary properties. 
The adopted grid in metallicity is Z=0.02, 0.016, 0.012, 0.008. 

With extinction:
\begin{equation}
\label{gen}
F_{\lambda}=\int_{t_f}^{t_i} \psi(t)\, B_{\lambda}(t,Z(t))\, 
f_{ext}(\lambda,t)\, {\rm d}t
\end{equation}
The wavelength dependence of $f_{ext}$ results from the assumed extinction 
model, as recalled below.

\subsection{Adopted extinction models.}

The extinction in galaxies depends on two factors : the spatial 
distribution of the dust and the properties of the dust grains, 
which we describe with an opacity coefficient $k_{\lambda}$. 
In Sects.\,\ref{parscreen} and \ref{cloudscreen}, we 
adopt simple models which assume the knowledge of the dust distribution 
and of $k_{\lambda}$. Opacity curves are
known to vary from one galaxy to the other, or even from
one line of sight to the other within the Milky Way. 
For the foreground extinction due to our own Galaxy, we use
the opacity  curve of Seaton (1979) at optical wavelengths,
and of Howarth (1983) in the UV and IR.
This curve is an average over many lines of sight towards Milky Way stars. 
Due to the presence of graphite grains, it shows a characteristic feature 
around 2200 \AA.
Because the dust in many galaxies has an extinction curve that lacks
a 2200 \AA\ bump, it has been argued that the Small Magellanic Cloud (SMC) 
opacity is more convenient for the modeling of galaxies' intrinsic extinction
(Gordon et al. 1997).
For the intrinsic extinction of our synthetic galaxies, we
adopt the SMC opacity curve of Pr\'evot et al. 
(1984) as listed by Gordon et al. (1997),
except where we explicitly allow it to vary.

In reality, the interstellar medium of a galaxy is 
not homogeneous, and is probably a mixture of different types of clouds 
(diffuse and compact) and different types of grains. Moreover, the 
light is scattered by the interstellar matter, so it is not possible 
to model the effective dust opacity in a simple way
(e.g. Witt \& Gordon 2000). This led
Calzetti et al. (1994, 2000) to derive an obscuration curve 
empirically from the integrated spectra of a sample of 
diverse starburst galaxies (Sect.\,\ref{empscreen}). 
In Sect.\,\ref{optdepth} we generalize this idea: it
is suggested to use the inversion technique to recover
both the wavelength dependence of the effective 
optical depth, and the time dependance of the SFR.

\subsubsection{Foreground dust screen.}
\label{parscreen}

In this simplest model, one has:
\begin{equation}
f_{ext}(\lambda,t)=e^{-\tau_{\lambda}(t)}=
\exp(-0.921\, k_{\lambda}\, E(B-V)(t)) 
\end{equation}
where $E(B-V)(t)$ is proportional to the column density of dust.
Stellar populations with different ages can be affected with different 
amounts of extinction, but all stars of a given age are affected identically.
As already mentioned, the values of $k_{\lambda}$ are those
of Pr\'evot et al. (1984). In this case, $\tau_{\lambda}$ is directly 
representative of the optical properties of dust grains on the line of sight 
towards stars. Representative plots of $k_{\lambda}$ can be found,
for instance, in the review of Calzetti (2001).

\subsubsection{Dust clouds in front of the stellar population.}
\label{cloudscreen}

In real galaxies, it is unlikely that all stars even of a given 
age see the same distribution of dust. Various mixed gas+stars models
and clumpy gas distribution models have been considered in the 
literature, to allow for this natural complexity. The model
considered here consists of dust clouds (or clumps)
distributed between the stars and the observer. The number of clumps on
the line of sight to a star of age $t$ obeys Poisson statistics\,:
$\overline{n}(t)$ describes the average number of clumps along
the line of sight. With
this model, a fraction $\exp(-\overline{n}(t))$ of all stars 
of age $t$ are seen without any obscuration. 

As demonstrated in Appendix\,\ref{dust_clouds.sec}, $f_{ext}$ has
the following analytical expression for the clumpy model\,:
\begin{equation}
f_{ext}(\lambda,t)=\exp\left\{-\overline{n}(t)
\left(1-e^{-\tau_{\lambda ,c}} \right)\right\}
=e^{-\tilde{\tau}_{\lambda}(t)}
\label{clumps.eq1}
\end{equation}
In this equation, $\tau_{\lambda ,c}$ is the optical depth for a single 
cloud:
\begin{equation}
\tau_{\lambda ,c}= 0.921\,  k_{\lambda}\,  E(B-V)_c
\label{clumps.eq2}
\end{equation}
$E(B-V)_c$ is the colour excess for one cloud. The resulting
extinction is described equivalently with an effective optical depth
$\tilde{\tau}_{\lambda}(t)$.

Note that when $\tau_{\lambda ,c}<<1$, $f_{ext}$ becomes
indistinguishable from a dust screen with total optical depth
$\overline{n}(t)\,\tau_{\lambda ,c}$, and when
$\tau_{\lambda ,c}>>1$ the attenuation becomes independent
of wavelength. In the former case, the attenuated spectra contain information
on the product $\overline{n}(t)\,\tau_{\lambda,c}$ but the two factors
cannot be separated; in the latter case, no signature of extinction
will appear in the spectrum (the dust would only be revealed by 
infrared emission).

\subsubsection{Empirical extinction model of Calzetti et al.}
\label{empscreen}

In the starburst extinction 
models of Calzetti et al. (1994, 2000), $f_{ext}(\lambda,t)$ 
is an unknown which can be determined directly from observations assuming
that all galaxies in the samples have similar star formation
histories.\\ 
Again, $f_{ext}$ is written:
\begin{equation}
f_{ext}(\lambda,t)=e^{-\tilde{\tau}_{\lambda}(t)}
\end{equation}
where :
\begin{equation}
\tilde{\tau}_{\lambda}(t) \propto \tilde{k}_{\lambda} E(B-V)(t)
\end{equation}
E(B-V) is the colour excess to the stellar continuum.
The empirical obscuration coefficient $\tilde{k}_{\lambda}$
results from the wavelength dependence of the grain
properties, as well as from the space distribution
of the dust, averaged over a typical starburst galaxy.
The proportionality constant in the above relation
is determined by Calzetti et al. (2000) from the energy budget of the galaxies,
which requires far-IR observations.

\subsubsection{Effective optical depth.}
\label{optdepth}

Inversion methods make it possible to explore whether
the available spectrophotometric data 
contain enough information to recover both
the time dependence of the SFR and the 
effective optical depth $\tilde{\tau}_{\lambda}$.
Thus, we may consider the effective optical depth, 
$\tilde{\tau}_{\lambda}$, 
as an unknown. In this paper, we then restrict ourselves
to the assumption of a constant $\tilde{\tau}_{\lambda}$
in time.

In fact, the opacity information lies principally in the emission lines.
The differential optical depth between wavelengths $\lambda_1$
and $\lambda_2$ is given by:
\begin{equation}
\ln\left(\frac{({\rm H}_{\lambda_1}/{\rm H}_{\lambda_2})_R}
{({\rm H}_{\lambda_1}/{\rm H}_{\lambda_2})_0} \right)=
\tilde{\tau}_{\lambda_2}-\tilde{\tau}_{\lambda_1}
\end{equation}
where ${\rm H}_{\lambda}$ is the emission in the line at
wavelength $\lambda$. The subscripts $R$ and $0$ correspond, 
respectively, to the reddened and the intrinsic emission line ratios.
In a similar fashion, the slope of the spectrum yields complementary 
information on the opacity distribution.
However, this approach does not provide the ratio of 
absolute to differential extinction.
If $\tilde{\tau}_{\lambda}$ is a solution, another 
possible solution is given by:
\begin{equation}
f'_{ext}(\lambda) = \exp(-\tilde{\tau}_{\lambda}+A)
\end{equation}
where $A$ is an arbitrary constant. The solution for the SFR 
then becomes:
\begin{equation}
\psi' (t)=\psi(t) \exp(-A)
\end{equation}
So, in the framework described in this subsection, one only obtains
information about relative fluctuations in the SFR.
In practice, one could force the curve $\tilde{\tau}_{\lambda}$ to vanish
when $\lambda$ tends towards infinity or use the far-infrared
thermal emission to determine the absolute extinction.\\

\section{The inverse method.}
\label{inv}

The determination of scalar functions (such as the SFR and the 
extinction law)  from observational data
is an underdetermined problem,
because the amount of observational data is only finite and thus
cannot provide the information for every detail of these functions.
Application of a straight inversion technique to Equation\,\ref{gen} could
be very sensitive to the noise in the data, and could
well give mathematically correct but unphysical results 
(Craig and Brown, 1986). The problem
must be regularized, which corresponds to a smoothing
operation (Twomey, 1977; Tikhonov \& Ars\'enine, 1976).

\subsection{A generalized least-squares approach.}

The inverse method we use comes from statistical techniques that have
been applied in geophysical analyses (Tarantola and Valette, 1982\,a,b;
Tarantola \& Nercessian 1984; Nercessian et al. 1984).
The method resembles Bayesian approaches in that {\em a priori}
ideas about the unknowns are used to regularize the inversion.

The conditional  probability density   $f_{\rm post}(M|D)$ for the
vector  $M$ of  the  unknown parameters,
given the observed data $D$ obeys:
\begin{equation}
f_{\rm post}(M|D) \propto {\cal L}(D|M)f_{\rm prior}(M)
\end{equation}
where $\cal L$ is the likelihood function and $f_{\rm prior}$ stands for the
a priori probability distribution for the model parameters.
$M$ will typically contain a sequence of values of the SFR at 
all ages, followed by the other adopted functional or discrete parameters
(e.g. the metallicity at all ages, the optical depth in the case
of a dust screen, the effective optical depth at all wavelengths
in the framework of Sect.\,\ref{optdepth}).
If we assume  that {\sl both }  the a priori probability and
the errors in the data are distributed as  Gaussian functions, 
and that there are no correlations between data and model parameters,
we can write :
\begin{eqnarray}
\lefteqn{ f_{\rm post}(M|D) \propto 
 \exp\bigg( \! -\mbox{$\frac{1}{2}$}(D-g(M))^{*}
 \cdot C_d^{-1} \cdot (D-g(M)) }    \nonumber\\
 & \qquad \qquad -\frac{1}{2}(M-M_0)^{*} \cdot C_0^{-1} \cdot
(M-M_0)\bigg)
\label{fpost}
\end{eqnarray}
$M_0$ is the mean of the prior (i.e. the a priori guess for the 
model parameters), and $C_0$ is the variance-covariance matrix
of the prior (see Sect.\,\ref{Co.sec}). 
The operator $g$ associates observable properties with 
model $M$: $D=g(M)$ if the model parameters are correct and there
is no noise in the data. In our case, Equation\,\ref{gen}
shows that $g$ provides the spectral energy distribution $F_{\lambda}$
through a set of equations of the form $F_{\lambda}=g_{\lambda}(\psi,Z,...)$. 
$C_d$ is the data variance-covariance matrix,
which describes the observational uncertainties. 
The superscript * refers to the adjoint operator.

The best estimate $M$ minimizes the quantity :
\begin{eqnarray}
\lefteqn{ \mbox{$\frac{1}{2}$}(D-g(M))^{*} \cdot C_d^{-1} \cdot (D-g(M)) }  
		\nonumber\\
& \qquad \qquad +  \frac{1}{2}(M-M_0)^{*} \cdot C_0^{-1} \cdot (M-M_0) 
\label{chi2gen.eq}
\end{eqnarray}
In the linear case, the minimum would be reached in one step  (Tarantola and
Valette, 1982\,a,b):
\begin{eqnarray}
\lefteqn{ M=M_0  + }    \nonumber \\
&  C_0  \cdot  G ^{*} \cdot
(C_d + G \cdot C_0 \cdot G^{*})^{-1} \cdot (D-g(M_0))
\label{min1}
\end{eqnarray}
$G$ is the matrix of partial derivatives of $g$. In our case however,
$g$ is not linear, and the minimum is reached iteratively:
\begin{eqnarray}
M_{[k+1]}& = & M_0 +     \nonumber \\
 & &  C_0  \cdot  G_{[k]} ^{*} \cdot
 (C_d + G_{[k]} \cdot C_0 \cdot G_{[k]}^{*})^{-1} \cdot    
			\nonumber \\
 & & (D + G_{[k]} \cdot (M_{[k]} -M_0)-g(M_{[k]}))
\label{min2}
\end{eqnarray}
$k$ counts the number of iterations and $G_{[k]}$ is
the matrix of partial Frechet derivatives at step $k$:
\begin{equation}
G_{[k]}=\left(\frac{\partial g}{\partial M}\right)_{[k]}
\end{equation}
In subsequent equations we shall abbreviate :
\begin{equation}
S_{[k]}:=C_d+G_{[k]} \cdot C_0 \cdot G^{*}_{[k]}
\end{equation}
To control the algorithm convergence at step $k$,
we test the stabilisation of $\chi^2$, with :
\begin{equation}
\chi^2_{[k]}=(D-g(M_{[k]}))^{*}
 \cdot C_d^{-1} \cdot (D-g(M_{[k]}))
\label{chi2.eq}
\end{equation}

\subsection{Choice of the model parameters and the prior $M_0$.}

\subsubsection{The a priori variance-covariance operator $C_0$}
\label{Co.sec}

The model parameters may include single value parameters as well
as functions (an example is described in detail in 
Appendix \ref{pardet}). For discrete parameters, $C_0$ is the matrix of the a
priori variances and a priori covariances between the parameters.
Large values of the variances 
must be chosen in the absence of any initial knowledge about the parameter 
values. Otherwise, the second term of expression\,\ref{chi2gen.eq} 
prevents the algorithm from moving away from a meaningless prior.
Let us define a temporal variable $u$ ($u=\log(t)$ or $u=t$ for instance;
see Appendix \ref{varchange}).
For an unknown function such as $\alpha(u)$ (i.e. the 
star formation history of Equation\,\ref{alphadef.eq}), 
defined at each time point,
$C_0$ incorporates a functional operator $C_{\alpha}$
(see Equation\,\ref{Co_equation}). Given the limited number of
data, we must assume some regularizing properties for $\alpha$.
\begin{equation}
\label{corsfr}
C_{\alpha}(u,u')=\sigma_{\alpha}(u) \sigma_{\alpha}(u')
\mbox{Cor}(u,u')
\end{equation}
The parameter $\sigma_{\alpha}$ is the prior
standard deviation. It controls
the acceptable amplitude of the deviations from the prior.
Cor($u,u'$) is the adopted autocorrelation function between two points of 
age $u$ and $u'$. We adopt:
\begin{equation}
\label{cor1}
\mbox{Cor}(u,u')=\exp\left(- \frac{(u-u')^2}{\xi_{\alpha}^2} \right)
\end{equation}

The parameter $\xi_{\alpha}$ is the prior correlation length.
It defines the resolution (in age) with which
one expects to recover the galaxy history functions.
A large value of $\sigma_{\alpha}$ will allow for more contrast in 
the function $\alpha(u)$ than a small one. 
On the other hand, if too large a value of $\xi_{\alpha}$ 
is chosen, the small (short timescale) details of the function $\alpha$ 
will disappear and the parameter fluctuations will become
too smoothed. The adequate choice of the two {\it a priori} parameters
($\xi_{\alpha}$ and $\sigma_{\alpha}$) is 
conditioned by the stability of the
inversion process, which itself is determined by the extent of the
available data and its signal-to-noise ratio (see also
Sect.\,\ref{kernel.sec}).

\subsection{Quality of the inversion.}

\subsubsection{The posterior variance-covariance matrix.}
\label{postvcov}

  Approximate internal errors on the 
  estimation of the  parameters can be computed once the minimization
  algorithm has converged. This is done by considering the local
  behaviour of the posterior probability density distribution 
  around the best solution (see also
  Reichardt et al. 2001, Moultaka \& Pelat 2000). From a second-order
  expansion of the posterior density distribution in this neighbourhood,
  one obtains:
  \begin{equation}
      C_{M} = C_0-C_0G^{*}S^{-1} G C_0
  \end{equation}

  Again, $C_{M}$ is a set of discrete values if the considered 
  parameters have discrete values. In the functional case, $C_{M}$ 
  is a set of functions of two variables. For instance, if the
  star formation history is one of the unknowns,
  $C_{M}$ incorporates terms of the form $C_{M,\alpha}(u,u')$.
  The diagonal terms such as $C_{M,\alpha}(u,u)$ measure
  how fast the exponent of Equation\,\ref{fpost} increases when 
  $\alpha$ is moved away from its best estimate.
  Note that in general this a posteriori variance is lower than the a priori 
  variance $\sigma_{\alpha}^2$ of Equation\,\ref{corsfr}. 
  The smaller the a posteriori variance is, compared to the a priori 
  variance, the more we have increased our knowledge of the 
  value of the parameter by using the available data.

\subsubsection{The resolving kernel and the mean index.}
\label{kernel.sec}

  Suppose that we knew the true model $M_{\rm true}$. Then the observed
  data would be (if the errors are negligible) :
  \begin{equation}
      D  = g(M_{\rm true})
  \end{equation}
  After the linearisation of $g$ near $M_{\rm true}$, 
  so that $G_{k} = G = g$,
  Equation\,\ref{min1} shows how the deviation of the
  true model parameters from a nearby initial guess would be
  degraded into the derived one :
  \begin{equation}
      \label{e:kernel}
         M-M_0 = C_0 G^{*} S^{-1} G(M_{\rm true}-M_0)
  \end{equation}
  The operator defined by
  \begin{equation}
       K := C_0 G^{*} S^{-1} G
  \end{equation}
  describes this degradation of information and is called
  the resolving kernel. Let's again,
  for illustration, focus on the part of $K$ relevant to 
  the star formation history $\alpha(u)$. Equation\,\ref{e:kernel} writes: 
  \begin{equation}
\alpha(u)-\alpha_0=\int K(u,u') (\alpha_{\rm true}(u')-
\alpha_0) {\rm d} u'
  \end{equation}
If the real star formation history deviated from a constant
by a delta function at age $u_{\delta}$, the shape of the
star formation history obtained by the algorithm would
be $K(u,u_{\delta})$. Because of the linearisation invoked, 
this response is only indicative. One of its major powers
is to draw attention to potential degeneracies, that may appear
in $K(u,u_{\delta})$ as very broad or multiple peaks.

  Another important and useful concept is a measure of the information
  present in the data. This is closely linked to the resolving kernel.
  Suppose that the parameter of interest is nearly constant 
  within the width of the kernel $K$. Then Equation\,\ref{e:kernel} gives
  \begin{equation}
      (M-M_0)(u)= (M_{\rm true}-M_0)_{\rm mean} \cdot
                          \int K(u,u') {\rm d}u'
  \end{equation}
  The integral is called the {\it mean index} $I(u)$
  \begin{equation}
      I(u) := \int K(u,u') {\rm d}u'
  \end{equation}
  and has the following meaning. If $I(u)$ has a very low value ($<<1$) 
  the model resulting from the minimization algorithm is expected to 
  lie close to the prior, whatever this prior is. 
  The quality of the $M$ estimate thus is poor. 
  But if $I(u)\simeq 1$ the model obtained from the algorithm is close 
  to the average true model. In other words, 
  only $I(u) \simeq 1$ ensures that the available data contained significant
  information on the estimated model parameter function. 
  In practice, a compromise
  has to be found. A mean index close to 1 can be obtained by 
  an increase in the smoothing length $\xi_{\alpha}$ of the a priori 
  variance-covariance operator. 
  However, increasing $\xi_{\alpha}$ reduces the resolution (in age).
  The adequate choice for $\xi_{\alpha}$ 
  is the smallest value that produces a mean 
  index close to 1.

\section{Determining SFR and reddening from mock spectra.}
\label{simul}

We have generated mock spectra from a given SFR ($\alpha(t)$), 
AMR ($Z(t)$), E(B-V) and average number of clouds ($\overline{n}(t)$) in 
order to test and illustrate the inverse procedure. 
The spectra extend through parts of the UV [1180-1680\,\AA], 
the optical [3690-6600\,\AA\ and 7000-9000\,\AA] 
and the near-IR [2.06-2.39\,$\mu$m],
as appropriate for the subsequent analysis of NGC\,7714 data 
(see Sect.\,\ref{ngc}). The spectral resolution matches that of the
models, which are based on the library of stellar spectra of
Lejeune et al. (1998; $\lambda/\Delta \lambda \simeq 100$).
The IMF of Scalo (1998) is adopted.

\begin{figure*}[!htb]
   \centering
   \includegraphics[angle=-90,width=0.95\textwidth]{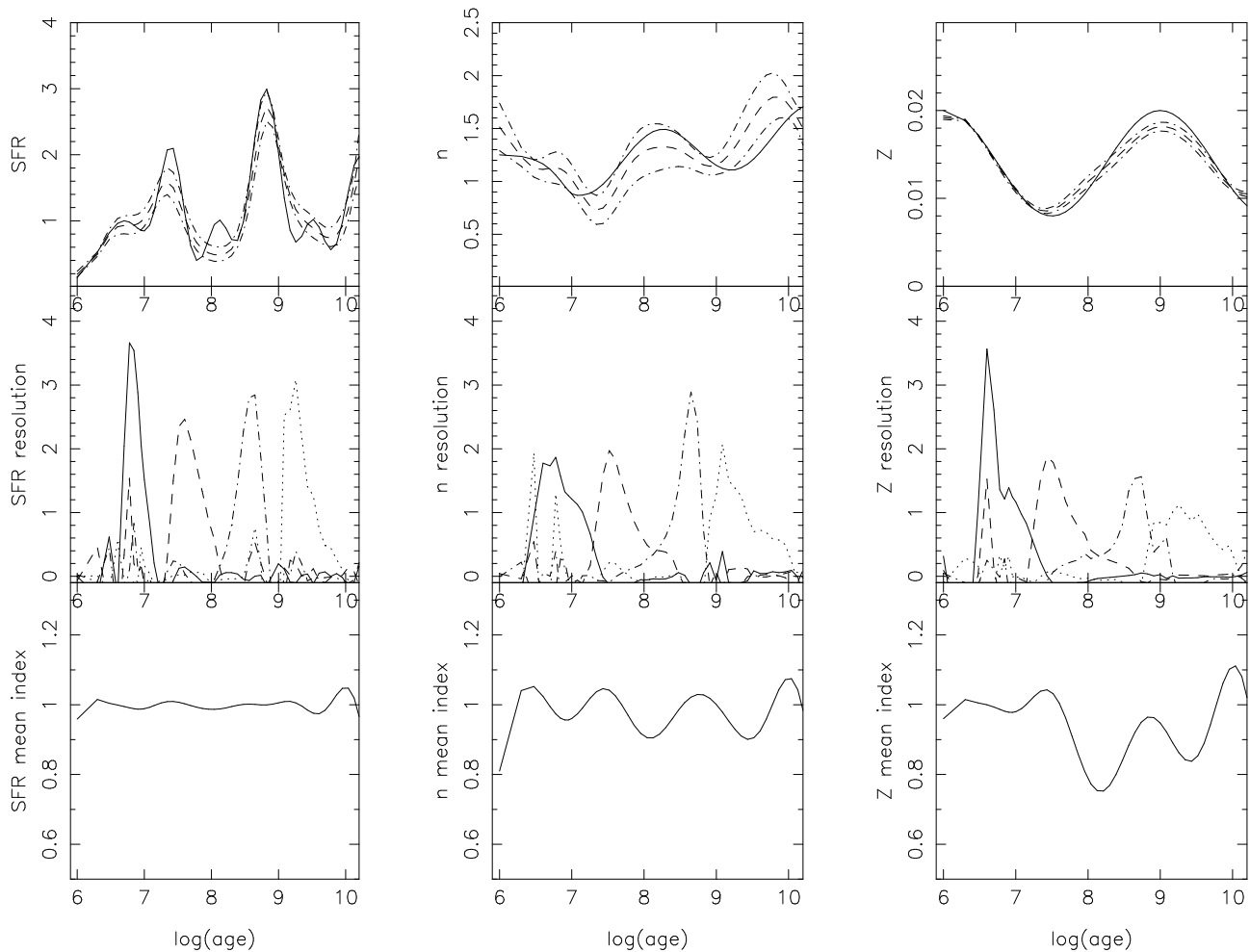}
      \caption{Inversion results from a spectrum with a signal-to-noise
      ratio of 500. The left column focuses on the
      the SFR, the middle column on the mean number of dust clouds
      on the line of sight, and the right column on the AMR.
      In the top panels, the solid lines describe the inputs used
      to construct the mock spectrum. The dashed lines are
      the best estimates from the noisy mock spectrum. The dot-dashed
      lines, and the middle and bottom panels allow us to asses
      the quality of the inversion. In the middle panels,
      the response functions $K(u,u_{\delta})$
      are shown for ages $u_{\delta}$ = log$(t)$ = 6.8 (solid),
      7.7 (dashed), 8.5 (dot-dashed) and 9.4 (dotted).
      See text (Sect.\,\ref{simul}) for details.
      }
         \label{fig1}
   \end{figure*}

\begin{figure*}[!htb]
 \centering
 \includegraphics[angle=-90,width=0.95\textwidth]{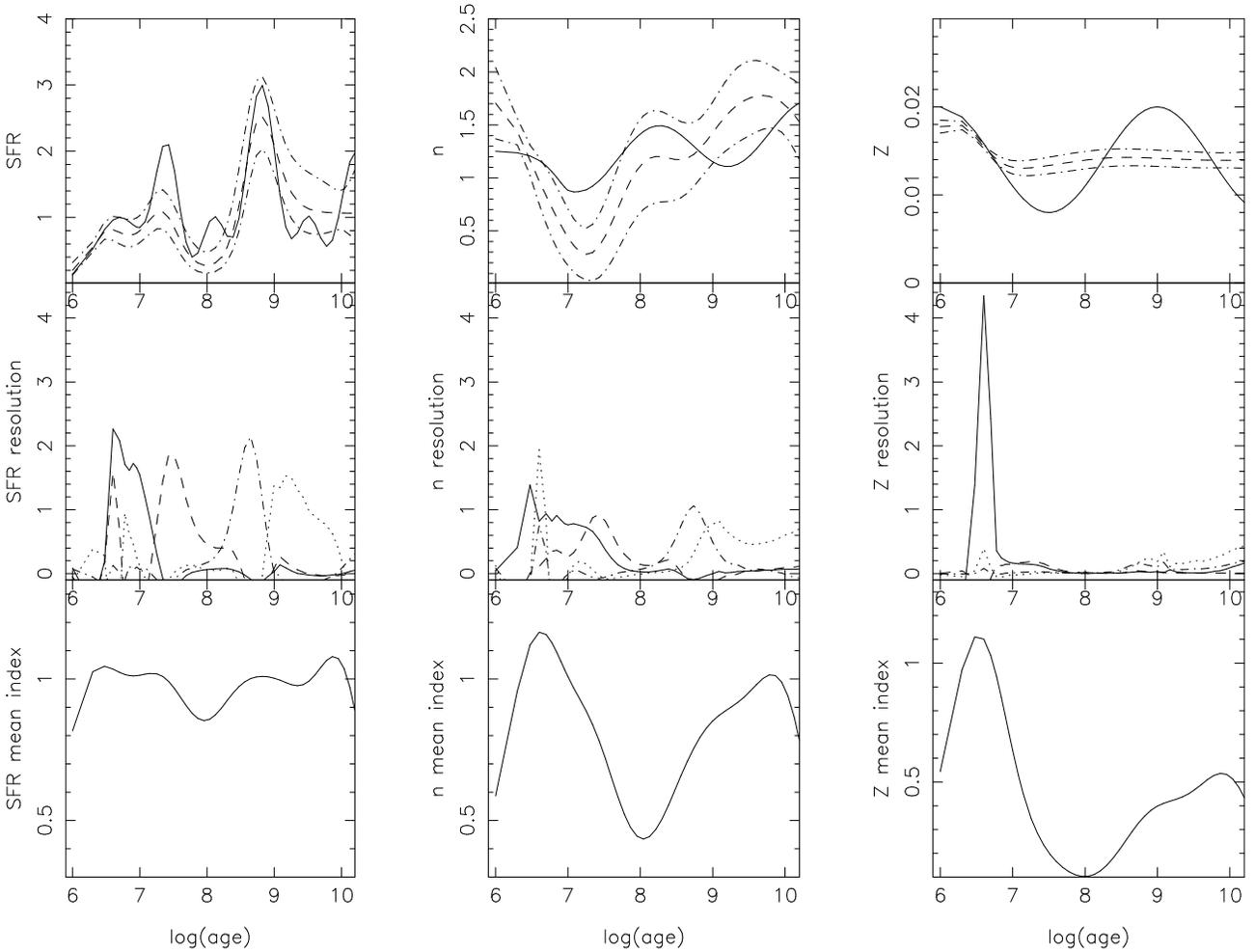}
 \caption{Same as previous figure from a mock spectrum with a 
signal-to-noise ratio equal to 25.}
 \label{fig2}
\end{figure*}

The noise added to the input 
spectra corresponds to a signal-to-noise ratio (S/N) equal 
respectively  to 500 (Fig. \ref{fig1}) 
and 25 (Fig. \ref{fig2}), and is assumed to be gaussian. 
Iterations were stopped when
$\chi^2$ was stable to one part in $10^4$ (typically 7 to 9 iterations).
The prior properties adopted for the figures are summarized
in Table\,\ref{testprior.tab}. 
The quality of the fits of the spectral energy distribution is evaluated
by means of a reduced $\chi^2$, $\chi^2_{\nu}$
(the $\chi^2$ of  Eq.\,\ref{chi2.eq} divided by the number of data points).
The results are not affected much by the prior provided 
that the reduced $\chi^2$ and the mean index are close to 1. 
Details of the implementation of the inversion method are provided
in Appendix\,\ref{pardet}.

\begin{table}
\caption[]{Priors for the inversions in Sect.\,\ref{simul}.}
\label{testprior.tab}
\begin{tabular}{|l|c|c|c|}
\hline
\multicolumn{4}{|c|}{\em Dust clouds with individual E(B-V)=0.2
	(Figs.\,\ref{fig1} and \ref{fig2})} \\ 
parameter & prior & prior $\sigma$ & prior $\xi$ \\ \hline
$\alpha(t)$     & 0     & 1    &0.3 - 0.5 [log($t$)]\\
$\overline{n}(t)$ & 1     & 1    & 0.3 - 0.5 [log($t$)] \\
E(B-V)        &0.5    & 0.25 & not applicable \\
Z$(t)$          & 0.014 & 0.001 & 0.5 [log($t$)] \\ \hline \hline
\multicolumn{4}{|c|}{\em Unknown effective optical depth (Fig.\,\ref{fig3}).}\\
parameter & prior & prior $\sigma$ & prior $\xi$ \\ \hline
$\alpha(t)$     & 0     & 0.5 &0.7 [log($t$)]\\
$\tilde{\tau}_{\lambda}$ & 1 & 3\,($\lambda\!<\!1\mu$m) & 
								1000\,\AA \\
	   &                            & 0.1\,($\lambda\!>\!1\mu$m) &         \\
\hline
\end{tabular}
\\
{\em Note\,:}\ $\alpha(t)$ is the natural logarithm of the 
SFR expressed in M$_{\odot}$\,yr$^{-1}$.
\end{table}

In Fig. \ref{fig1} the results obtained with S/N=500 are shown.
The input parameters are drawn in full lines in the top panels. 
The input value of E(B-V) for individual clouds is 0.2.
The dashed lines show the best estimates of these parameters
resulting from the inversion. They led to a $\chi^2_{\nu}$ of 1.028.
The dot-dashed lines show the
standard deviations from these best estimates, computed
from the diagonal terms of the posterior variance-covariance
matrix of Sect.\,\ref{postvcov}. For E(B-V), the resulting
estimate is 0.194, and the posterior standard deviation
is 0.03.

The global features in the 
SFR are recovered well, whereas the smallest 
are not. The posterior resolution can be examined in the
middle panels. They display the  
response of the inversion algorithm 
to delta-function inputs in SFR$(t)$, $\overline{n}(t)$ and
$Z(t)$, located at ages log$(t)$ = 6.8, 7.7,
8.5 and 9.4. The computation is based
on the resolving kernel $K$ of Sect.\,\ref{kernel.sec}. It is seen
that the resolution compatible with the mock data
is about 0.3-0.6 in log(age). The mean index for the 
SFR is uniformly close to 1 over the age range. Together
with the fact that the posterior standard deviation is
much smaller than the prior value, this tells us
the SFR is well constrained.

The average number of clumps and the AMR are not resolved as well as 
the SFR. The reddening of individual clouds, E(B-V), has 
been recovered well. Note that its value is compatible with the
recovery of $\overline{n}(t)$ (see end of Sect.\,\ref{parscreen}). 
If E(B-V) had been found close to 0 a degeneracy
with $\overline{n}(t)$ would have been suspected; if it had been very large
both its own value and $\overline{n}(t)$ would have been very uncertain.

Despite the high signal to noise ratio, the mean index 
associated with the AMR begins to weaken. Note that the smallest
values of the mean index tend to appear at ages where the 
star formation rate is low: it is not surprising that little
information on the metallicity of the corresponding stars
can be found in the data.

Figure \ref{fig2} shows how much information can be recovered when
the available data has S/N=25. Only the global trends are recovered
for the SFR and the average number of clumps. For the AMR, the
resolution and the mean index fall down after log(age)=7 (10 Myr) :
the recovered AMR corresponds to the prior. Searching for the
AMR should be avoided with this type of data.

\begin{figure}
   \centering
   \includegraphics[angle=0,width=0.45\textwidth]{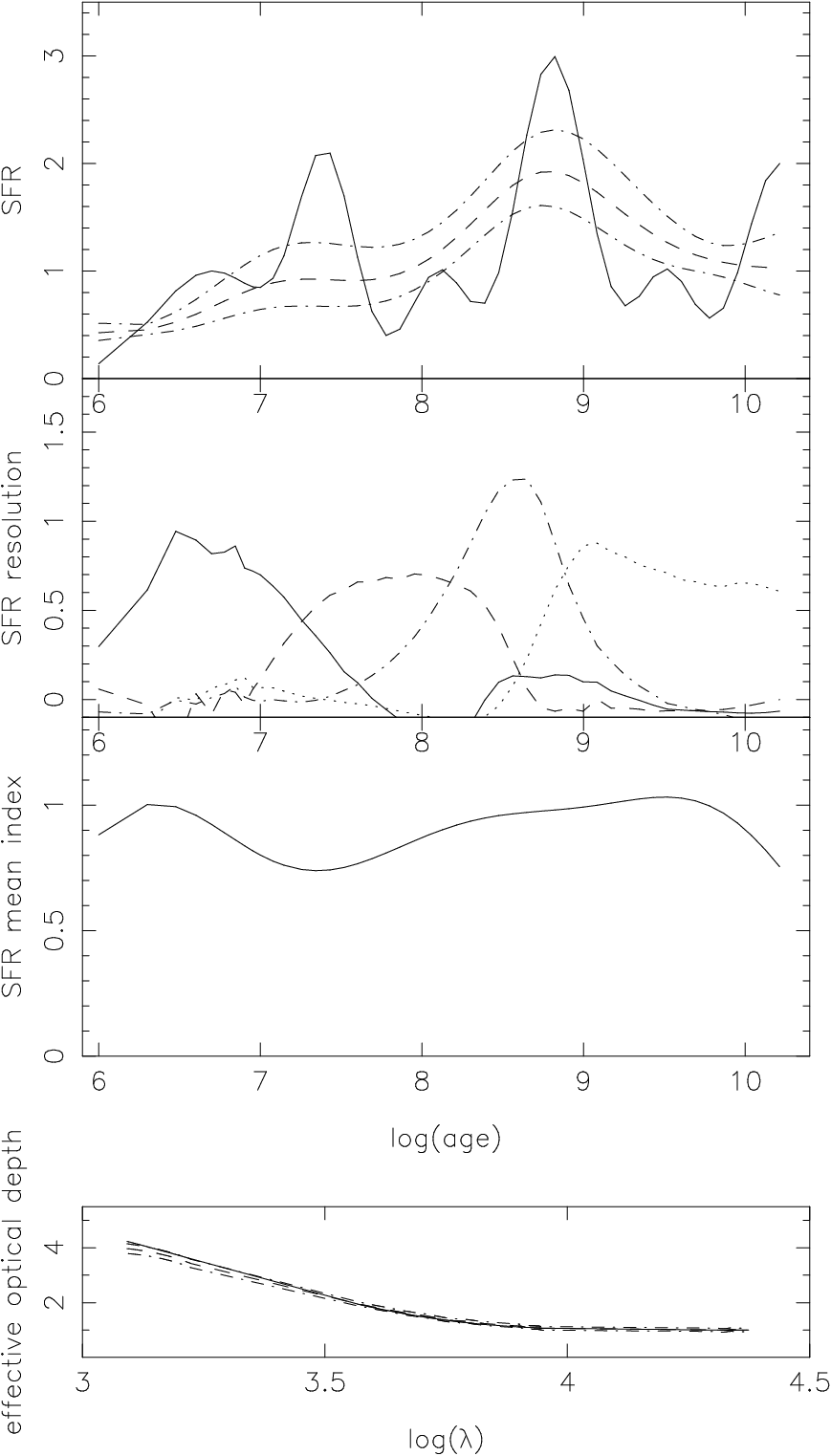}
      \caption{Inversion results from a spectrum with a S/N = 25, when
 the wavelength dependence of the effective optical depth  is also
 considered as an unknown. Metallicity is assumed to be known. 
 The first three panels correspond to the
 first column of Figs.\,\ref{fig1} and \ref{fig2}.
 The bottom panel shows the assumed optical depth (full line),
 the recovered one (dashed line) and the internal error (dot-dashed).
  }
         \label{fig3}
   \end{figure}

In Fig. \ref{fig3}, we test how well the method is able to reconstruct
the SFR and the effective optical depth $\tilde{\tau}_{\lambda}$
from simulated spectra with S/N=25 (see Appendix\,\ref{opa}). 
Here, we do not consider the AMR
as an unknown, because at this level of noise
and spectral resolution, the information present in the
data is too poor. As
mentioned in Sect.\,\ref{optdepth}, we assume $\tilde{\tau}_{\lambda}$
is the same for populations of all ages. The three
top panels show respectively the SFR, the resolution in age and the
SFR mean index as in previous figures. Only a large smoothing
length $\xi_{\alpha}$ gives a decent mean index: the age resolution
on the SFR is poor. 

The input and output effective
optical depths are presented in the bottom panel.
The mean index for $\tilde{\tau}_{\lambda}$ takes values close
to 1 in the UV, then drops progressively to poor values, 
of order 0.5 at 1\,$\mu$m and 0.1 around 2\,$\mu$m. The low index 
values at long wavelengths are due essentially to the  
degeneracy between the absolute values of the 
effective obscuration and of the SFR (see Sect.\,\ref{optdepth}):
with a small prior standard deviation, the prior sets
the result (with larger prior standard deviations, the algorithm
doesn't converge).
In Fig.\,\ref{fig3} we have forced the near-IR match of 
$\tilde{\tau}_{\lambda}$, so input and output SFRs are comparable.
As the UV emission can only come from young stars, 
the UV slope provides a strong
constraint on that part of the reddening law. The emission
of the young component at longer wavelengths does not by 
itself match the data. The optical and near-IR fluxes
directly provide a minimum contribution of older stars and
the main constraints on their age distribution. 
Freedom in the extinction law acts as a correction to improve the fit.
A relatively long prior correlation length for $\tilde{\tau}_{\lambda}$
(1000\,\AA ) avoids excessive freedom, that would lead to the fitting
of the noise in the data.

In view of the limited resolution, spectral coverage and signal-to-noise
ratio assumed here, the results of Fig.\,\ref{fig3} are very satisfactory
and the method is promising.

\section{An example : NGC 7714.}
\label{ngc}

NGC 7714 is a well studied interacting spiral galaxy which hosts a starburst in 
its nucleus (Arp 1966, Weedman et al. 1981, Calzetti 1997). 
The peaked ground based morphology of the starburst, the
small inclination of the galaxy disk seen from our perspective, 
and the relatively low extinction suggested by the ultraviolet (UV) 
brightness (Markarian \& Lipovetskij 1974), 
motivated detailed studies of this apparently simple system. 
As a result of high resolution imaging and 
spectroscopy, the simplest models for the nucleus
had to be successively excluded (Gonz\'alez-Delgado et al. 1995, 1999,
and references therein). The central $\sim 330$\,pc of the
galaxy were found to contain a variety of stellar populations, ranging
from very young ($\sim 5$ Myr) to old (the population of the 
underlying spiral). Simple dust screen extinction models, that were
initially compatible with the scarce available data (Puxley \& Brand, 1994),
were found to be inappropriate (Lan\c{c}on et al. 2001, hereafter
LGLG01).

The study of the central $\sim 330$ pc of the galaxy by LGLG01
was based on a ``manual" exploration of a set of model star formation
histories. The SFR was assumed to be a combination of a maximum of four
components, each component being represented either by a standard spiral
galaxy SFR, an instantaneous burst, an episode of constant star formation
or an exponentially decreasing star formation episode. The exploration
of this parameter space was guided by preliminary studies in individual
wavelength ranges. Such an exploration method is tedious and cannot
be exhaustive. A chance remains that the best fitting model may be missed.
On the other hand, this exploration has provided a range of constraints
that are considered robust because they are common to most of the
satisfactory model adjustments to the data. 
Among those were the following.

(i) The nucleus of NGC\,7714 has been forming stars 
off and on over the past several hundred millions of years at an average
rate of the order of 1\,M$_{\odot}$\,yr$^{-1}$, with a brief enhancement
of a factor of a few about 5\,Myr ago.

(ii) The extinction even in the central 300\,pc of NGC\,7714 is
inhomogeneous; for instance, most of the UV luminosity of this area is 
due to an obscuration-free line of sight towards a young cluster
which does not coincide with the maximum Brackett\,$\gamma$ (Br$\gamma$)
emission. 

(iii) Assuming a constant IMF in time, 
the recent star formation episodes have enhanced the stellar 
mass in the central area by at least 10\,\%, and more likely 
by 25\,\%. 

(iv) Most of the satisfactory models implied that the level of star 
formation had increased already several 100\,Myr ago. This timescale
called for confirmation, as dynamical modeling of the interacting system
indicates that only about 100\,Myr have elapsed since closest approach between 
NGC\,7714 and NGC\,7715 (Smith \& Wallin 1992, Smith et al. 1997, 
Struck \& Smith 2002). 

The existence of multiwavelength data and previous detailed investigations
of plausible models for the central regions of NGC\,7714 make this
galaxy a target of choice for the application of our automatic inversion method.
The available data are described in detail in LGLG01. The spectral
coverage is as in Sect.\,\ref{simul}.
We have degraded the resolution of the 
observations to match the
wavelength sampling of the models. 
A correction for foreground extinction
due to the Milky Way was also applied (E(B-V)=0.08). 
The signal-to-noise ratio varies along the spectrum 
with a typical value of 25. The gap in 
the optical part of the spectrum is due to the removal of data
of poorer quality (due to telluric absorption).

\subsection{Estimation of the SFR with simple extinction models.}
\label{extmod.sec}

In this section, we apply the three extinction models presented previously 
in Sect.\,2.2: a screen of SMC-type dust, clouds of SMC-type dust, 
and the empirical attenuation law of Calzetti et al. (2000).
The priors are listed in Table\,\ref{7714prior.tab}
and discussed in Sect.\,\ref{prior_disc.sec}.

\begin{table}
\caption[]{Priors for the inversions in Sect.\,\ref{ngc}.}
\label{7714prior.tab}
\begin{tabular}{|l|c|c|c|}
\hline
\multicolumn{4}{|c|}{\em Inversions of Sect.\,\ref{extmod.sec},
\ref{metal.sec} and \ref{newbasis.sec}.} \\
parameter           & prior & prior $\sigma$ & prior $\xi$ \\ \hline
$\alpha(t)$       & -1.2  & 0.4  & 0.5 [log($t$)]\\
\multicolumn{3}{|l|}{\em ~~ SMC or Calzetti-type dust screen:} & 
 \rule[0pt]{0pt}{12pt} \\
{\small E(B-V)(t)}   & 0.2  & 0.2  & 0.5 [log($t$)] \\
\multicolumn{2}{|l|}{\em ~~ Dust clouds:}  & & 
 \rule[0pt]{0pt}{12pt} \\
$\overline{n}(t)$ & 2     & 2    & 0.5 [log($t$)] \\
E(B-V)$_{c}$      &0.2    & 0.2  & none \\ 
\hline \hline
\multicolumn{4}{|c|}{\em Inversions of Sect.\,\ref{freelaw.sec}.}\\
%
parameter & prior & prior $\sigma$ & prior $\xi$ \\ \hline
$\alpha(t)$       & -1.2  & 0.8  & 0.6 [log($t$)]\\
ln$(\tilde{\tau}_{\lambda})$ & {\small 0\,($\lambda\!<\!1\mu$m)}  
			& {\small 1.5\,($\lambda\!<\!1\mu$m)} & 3000\,\AA \\
           &  {\small -2\,($\lambda\!>\!1\mu$m)}  &
		{\small 0.1\,($\lambda\!>\!1\mu$m)} &   \\
\hline
\end{tabular}
\\
{\em Note\,:}\ $\alpha(t)$ is the natural logarithm of the
SFR expressed in M$_{\odot}$\,yr$^{-1}$.

\end{table}


\begin{table}
\begin{center}
\caption[]{$\chi^2_{\nu}$ values for different extinction models and different
IMF slopes.}
\label{chiext}
\begin{tabular}{|p{2.cm}|r|r|}
\hline
model & Scalo IMF & Salpeter IMF \\
\hline
dust screen & 2.5 & 2.3 \\
\hline
dust clouds & 2.3 & 2.0 \\
\hline
Calzetti & 2.1 & 1.9 \\
\hline
\end{tabular}
\end{center}
\end{table}

\begin{figure}[!hbt]
\centering
%
\includegraphics[angle=0,width=0.45\textwidth]{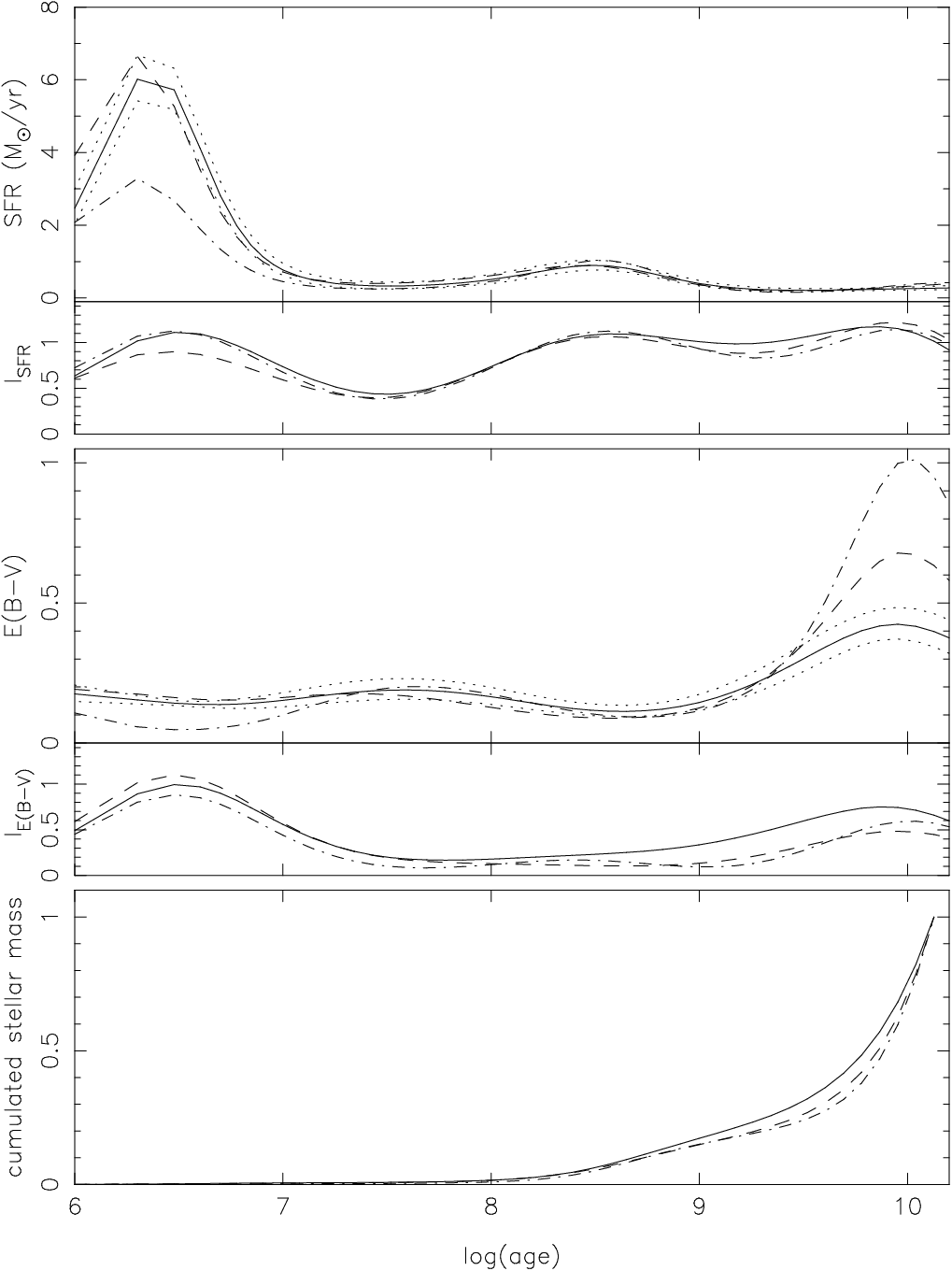}
%
\caption{Properties obtained for the nucleus of NGC 7714,
assuming a Salpeter IMF and solar metallicity (Z=0.02).
The top panel shows the star formation histories.
The middle panel shows the differential extinction  and the bottom
panel the cumulative contributions of stars of various 
ages to the total stellar mass. 
Solid: Calzetti dust (the errors given by the posterior standard
deviation are indicated with dotted lines in the top panel); 
dot-dashed: dust screen; dashed: dust clouds.
}
\label{figext1}
\end{figure}

The main results are plotted in Fig.\,\ref{figext1}.
The derived star formation histories and extinctions are in
good global agreement with the previous study of LGLG01.
A bimodal SFR is found, with two peaks respectively 
centered at log(age)=6.4 (2.5\,Myr) and 8.5 (300\,Myr). 
The average level of the SFR over the last few 10$^8$\,yr
is of the order of 1\,M$_{\odot}$\,yr$^{-1}$. 
Stars younger than $10^9$\,yr contribute about 
20\,\% of the total stellar mass that has ever been produced.
The most recent episode of activity however, despite its high star formation
rate, only added a few percent to the stellar mass.

The posterior standard deviation and mean index shows
that the SFR is well constrained, except around log(age)=7.5.
The stars at these ages contribute little to the light, compared
to the youngest burst in the UV, or to the
bulk of the somewhat older intermediate age stars at
optical wavelengths. 
The amount of obscuration is constrained
tightly for the youngest stars, 
and reasonably well for the oldest,
but the mean index is poor at intermediate ages. 

In the case of the cloudy dust model, $\overline{n}(t)$ and the 
reddening per cloud are searched for. $E(B-V)_c=0.17$ is found for the
individual clouds (with a posterior standard deviation of 0.03). 
$\overline{n}(t)$ varies between 1 and 2 at young and intermediate ages, 
and increases to about 6 for the old populations. $\overline{n}(t)$ 
and $E(B-V)_c$ are combined using 
Equation\,\ref{clumps.eq1} and \ref{clumps.eq2} 
to provide the final colour excess plotted in Fig.\,\ref{figext1}.
Note that this value is systematically smaller than the simple
product $\overline{n}(t)\,E(B-V)_c$.

\begin{figure}[]
\centering
\includegraphics[angle=0,width=0.45\textwidth]{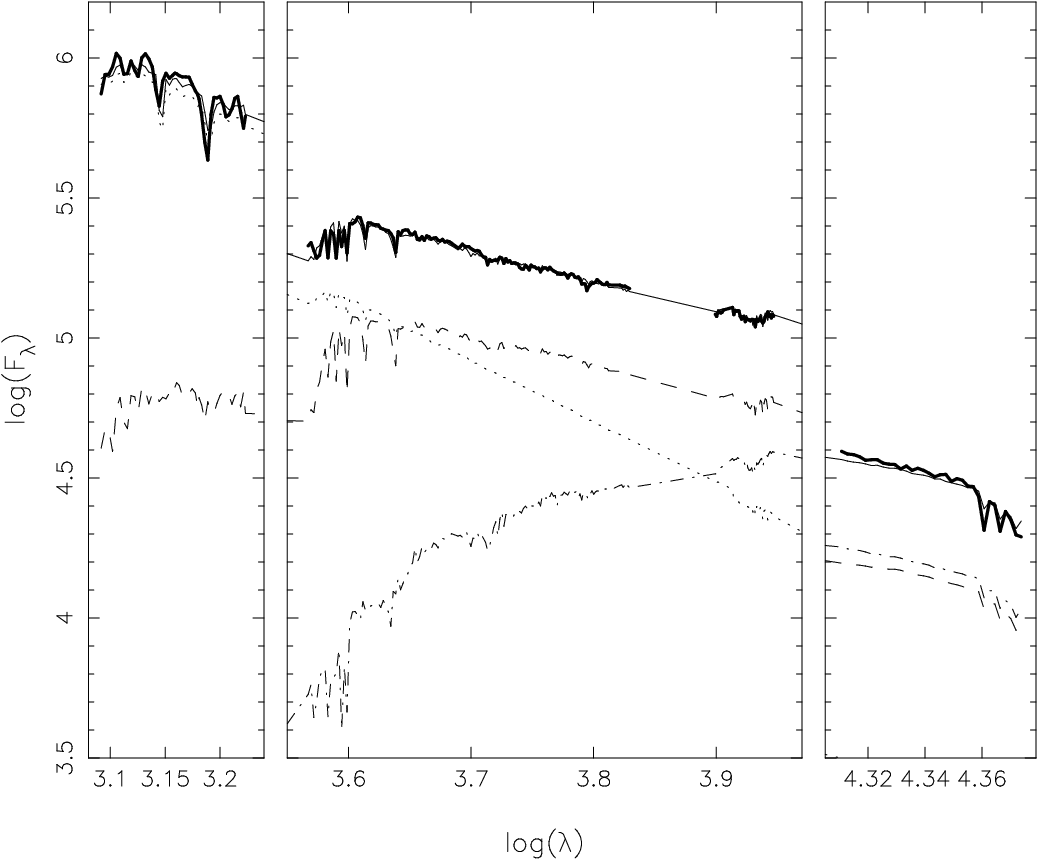}
\caption{Adjustment of the NGC\,7714 data obtained with the dust 
description of Calzetti. Thick line: data (gaussian fits to the
emission lines have been subtracted before resampling). Thin solid line:
model SED (interpolated through regions with no data). Also shown are the
respective contributions of stars in the following age ranges:
$<10^7$\,yr (dotted), $10^7-10^9$\,yr (dashed), $>10^9$\,yr (dot-dashed).
}
\label{Calzfit.fig}
\end{figure}

The dependence of the colour excess on stellar age is
consistent with the results of LGLG01. Relatively large
values are found at old ages. This trend is necessary in order 
to reach the near-IR flux level without compromising 
the fit at optical wavelengths (Fig.\,\ref{Calzfit.fig};
see however Sect.\,\ref{prior_disc.sec}). 

In the UV, the SMC opacity curve is the steepest of 
the reddening laws we have considered. With that law, the 
blue slope of the UV emission of NGC\,7714 
sets the strongest limitation on the amount of 
dust towards young stars. Consequently, a smaller 
SFR is sufficient to produce the observed level of UV flux. 
The shape of the opacity curves also
explains the differences at old ages: the different E(B-V)
of Fig.\,\ref{figext1} produce old components of very similar
shape and contributions in plots such as Fig.\,\ref{Calzfit.fig}.

The reduced $\chi^2$ values in Table\,\ref{chiext}
show that the effect of the IMF on the quality of the
best fits is small, the Salpeter IMF being favoured.
The clumpy dust model and the attenuation law of Calzetti et al.
produce a slightly better agreement with the observed SED than a
simple SMC dust screen.

An important caveat for all the results of this section
is that the Br$\gamma$ emission line is
underestimated by a factor of two by the models,
and this even though we do not apply stronger
extinction to the emission lines than to the stellar
continua (i.e. we deviate from the prescription of Calzetti et al.
2000 in this point).
The missing Br$\gamma$ emission corresponds to a nebular emission excited
by a young population that is not found by the inversion
algorithm when the wavelength dependence of the extinction
is one of the laws considered here. The young population is likely 
to be underestimated. We note that the same problem
was faced by LGLG01 when they sought to adjust only the continuum,
and led them to invoke additional highly obscured young stars.
A similar {\em ad hoc} fix would also be successful here.
As a partial answer to the lack of Br$\gamma$ photons,
the age found by automated inversion for the most recent starburst 
is younger than the 5\,Myr of LGLG01.

\subsection{Discussion of the inversion assumptions}
\label{prior_disc.sec}
It is important to 
analyse the effects of the adopted priors.
As expected, the numerical values of the priors given in 
Table\,\ref{7714prior.tab} are of no consequence where the mean index 
is close to 1, as long as they remain in a range in gross agreement 
with the flux level of the data, i.e. within reach of the iteration
procedure given the adopted prior standard deviations. 
In practice, reasonable starting points are obtained after 
a small number of initially random trials which 
show into which direction the algorithm pulls.

Increasing the prior standard
deviations or reducing the correlation lengths significantly
from the values in the table results either in loss of
convergence, or in unacceptably small values of the mean
indices, or in large posterior variances:
the information in the data becomes insufficient.

\begin{figure}[]
\centering
\includegraphics[angle=0,width=0.45\textwidth]{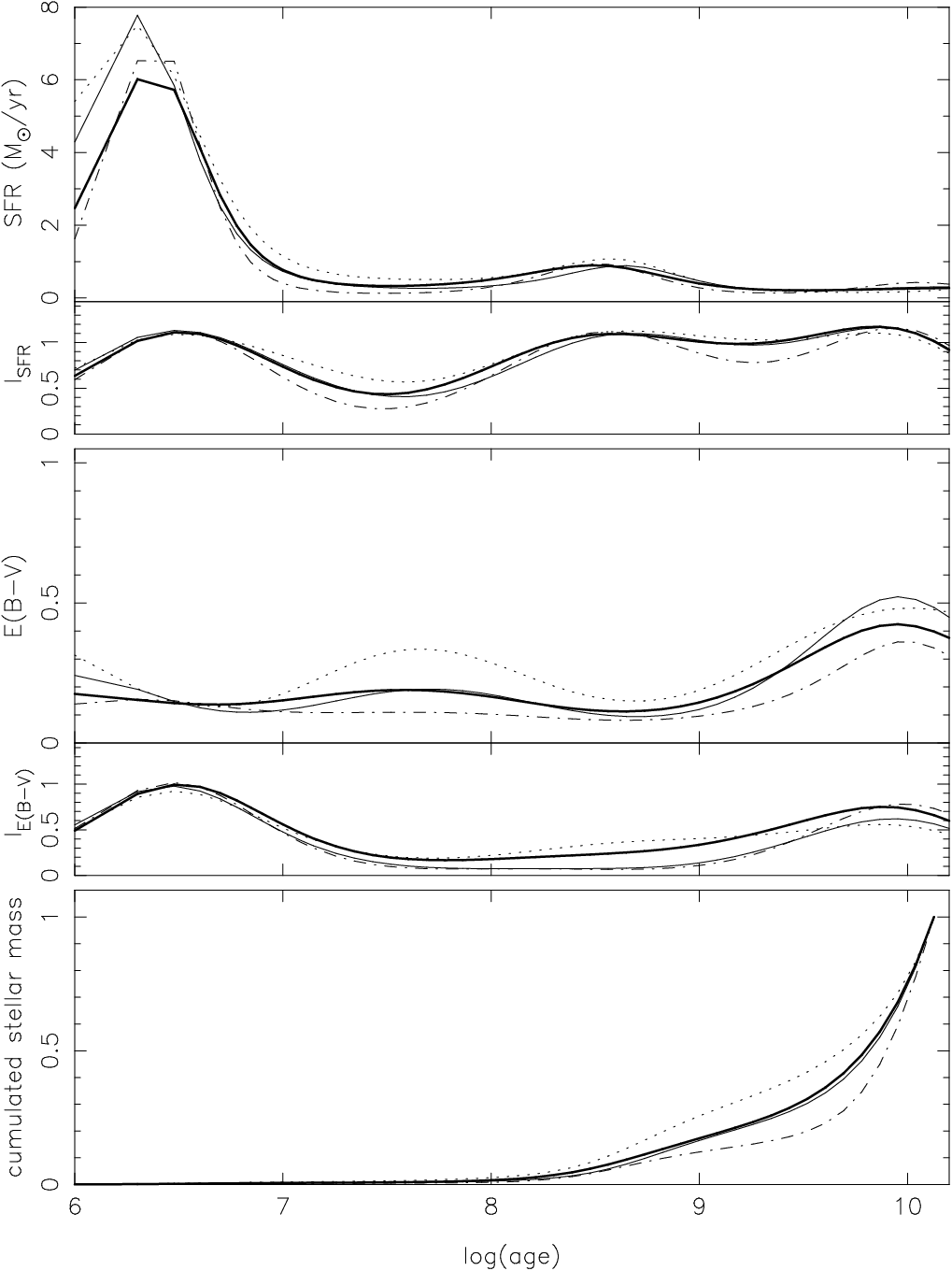}
\caption{Effects of the prior assumptions on the estimated parameters,
in the case of Calzetti-type dust.
The thick lines are the results of Fig.\,\ref{figext1}.
The thin solid line is the result of double iteration (see text).
The two other lines are obtained with priors designed to pull 
the SFR towards low (dot-dashed) and high (dotted) values at intermediate
ages (i.e. where the mean index of E(B-V) is low).
The values of $\chi^2_{\nu}$ remain between 1.85 (double iteration) and
1.94. 
}
\label{prior_effects.fig}
\end{figure}

As it is, the mean index of the extinction is low at 
intermediate ages. Nevertheless, the results are found
to be quite stable against changes in the priors. In 
Fig.\,\ref{prior_effects.fig}, extreme results are plotted
in the illustrative case of Calzetti-type dust.
They were obtained by modifying the initial guesses of both
the SFR and the extinction in order to pull towards higher or lower
resulting SFRs. The shape of the SFR appears to be robust,
although the mass ratio of old to relatively
young stars (which is sensitive to small changes in the actual 
value of the early SFR) appears to be uncertain to
within a factor of two.

Also plotted in Fig.\,\ref{prior_effects.fig} is the result obtained
when the outputs of Sect.\,\ref{extmod.sec} are used as priors
for a new iteration. This step provides a good test
of the initial convergence. The changes are satisfactorily small. 
After successfully performing the test in several cases, we
have stopped applying it systematically.

Finally, we have neglected until now the uncertainties 
due to possible aperture mismatch between the UV, optical
and near-IR observations. LGLG01 estimate that errors in the 
relative flux levels of the three spectral segments are below 15\,\%.
Although it is possible to build
this uncertainty directly into the inversion algorithm,
we have more conservatively run inversions on a series
of modified data. Factors between 0.85 and 1.15 were applied
to the UV and near-IR parts of the spectrum. The resulting 
fits are poor ($\chi^2_{\nu}>2.5$) when the near-IR flux is increased
above the reference value. The best fits ($\chi^2_{\nu}=1.8$) are obtained
with a reduced near-IR flux. These solutions are appealing
because they don't require higher extinctions for the old populations
than for the intermediate age ones. The earliest star formation rates
are correspondingly smaller, and the intermediate age and young populations
becomes more important in terms of total stellar mass. 
However, all the tests run here provide results within the global
envelope of the results already discussed above or in the previous
section. 

None of our attempts have a provided a solution
in which the increase in the SFR responsible for
the large intermediate age population occured less
than about 300\,Myr ago.

\subsection{Discussion of star formation history}
\label{sfr_disc.sec}

Struck \& Smith (2002) discuss possible causes for
a star formation episode that would have started 300 Myr or longer ago,
but favour none in particular.
The strength and extent of the stellar ring of NGC\,7714
and of the tidal tails of the system suggest that the
galaxies are observed shortly after closest approach and well before
their probable merger (Toomre \& Toomre 1972,
Barnes \& Hernquist 1992, Gerber \& Lamb 1994).
What ``shortly after" precisely means remains dependent on the
model. Struck \& Smith (2002) caution that modest changes of orbital
inclination or distance of closest approach can significantly
change the structure of the disk waves, and thus the central SFR
in NGC\,7714. At the same time, they admit a variety of imperfections in their
final choices (for instance, they attribute the lack of a strong
stellar ring in their hydrodynamical simulation to
a slightly overestimated impact parameter). Furthermore,
the timescale of the formation and disruption of tidal features
depends on impact speed (Dubinski et al. 1999).
Although the choice of initially parabolic galaxy orbits made
by Struck \& Smith is reasonable,
the previous histories of the galaxies, or galaxy halo structures
different from those adopted may have led to a slower
encounter with longer timescales.

Various dynamical or hydrodynamical simulations
of mergers produce double star formation episodes over timescales
of the order of 10$^9$ yr, but the second one usually corresponds
to the final merger, which is not appropriate for our case study
(Mihos \& Hernquist 1994, Bekki 1998, Gerritsen \& Icke 1999).
However, some models predict that each of these SF episodes, especially
the first one, will last for several 100\,Myr (Bekki 1998). If
these models incorporated feedback mechanisms associated with star formation,
such as local heating of the ISM that partially prevents star formation,
the predicted SFR would be much more irregular (Gerritsen \& Icke 1999)
and might describe what happened in NGC\,7714.

From the simulations available in the literature,
our conclusion is that the last interaction with NGC\,7715 
might explain SF timescales of up to about 200 Myr, 
while longer SF timescales call for a previous event. A
previous passage of the companion galaxy and related (or
unrelated) bar instabilities, are options worth considering
(Smith \& Wallin 1992, Friedli \& Benz 1995).

\subsection{The influence of the assumed metallicity.}
\label{metal.sec}

Section\,\ref{extmod.sec} considers a solar metallicity for NGC 7714. 
Here we investigate the effect of the assumed metallicity on the 
estimated SFR. 

\begin{figure}[!t]
\centering
\includegraphics[angle=0,width=0.45\textwidth]{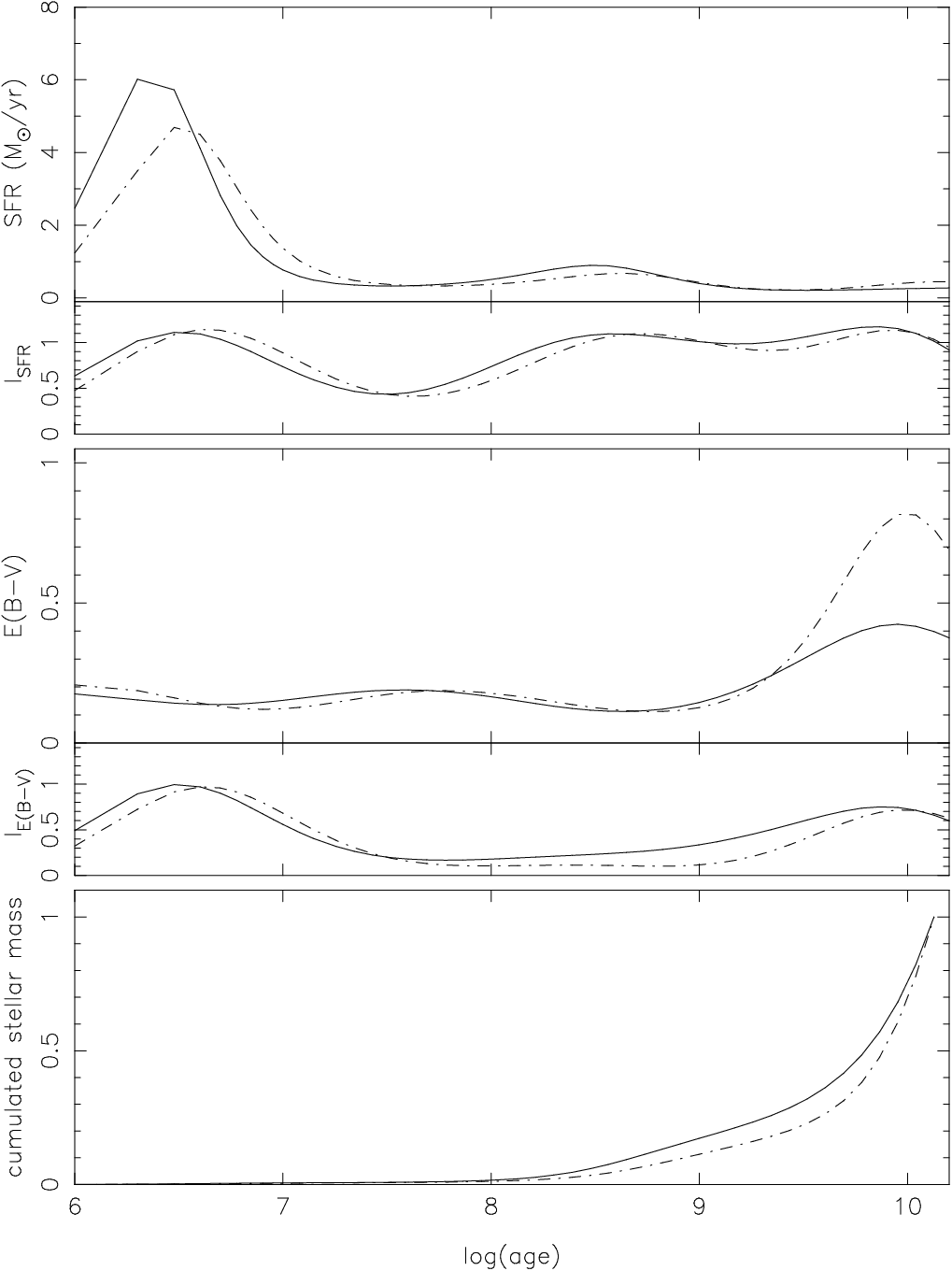}
\caption{Determination of the NGC 7714 star formation rate for 
two different metallicities, assuming a Salpeter IMF and
the attenuation law of Calzetti et al. (2000). 
Solid line: Z=0.02; dashed line: Z=0.008. 
The layout is as in Fig.\,\ref{figext1}. 
}
\label{met}
\end{figure}

Fig. \ref{met} compares the SFRs obtained 
for Z=0.02 and Z=0.008. Due to the smaller opacities
of metal poor stellar atmospheres, a stellar 
population of a given age is intrinsically bluer
at Z=0.008 than at Z=0.02. To reproduce the
spectrum with a given effective extinction law, 
a shift of the star formation episodes to somewhat larger
ages is therefore required at Z=0.008. Even the oldest 
stars are not red enough to explain the near-IR flux,
and enhanced reddening towards the old populations results.
The resulting absorption is compensated with a higher
early star formation rate, and the predominance of old
stars in the mass budget is reinforced.

The larger width of the recent star formation episode compensates 
for the lower value of the maximum SFR, and the total
amount of UV light produced is similar to the Z=0.02
case.

\subsection{Using a new basis of model spectra.}
\label{newbasis.sec}

Spectrophotometric properties of stellar populations with
complex star formation histories depend critically on the 
properties of the building blocks of evolutionary models, the
so-called single stellar population models that define the
basis B$_{\lambda}$. In addition to the basis of the previous
sections, we constructed a second set using the new grid of
Mouhcine (2001) and Mouhcine \& Lan\c{c}on (2002). In this grid
particular care was taken to model the near-infrared
properties of intermediate age stellar populations.
The major difference relative to the inputs of {\sc P\'egase}
lies in the modeling of the Thermally
Pulsing Asymptotic Giant Branch (TP-AGB). TP-AGB stars dominate
the near-infrared light of intermediate-age (0.1\,-2.\,Gyr)
stellar populations. 
As a result of the interplay
between mass loss, the third dredge-up process and, most
importantly, envelope burning, the new models 
predict that the TP-AGB stars with the longest lifetime
are born from stars with main sequence lifetimes of
0.8-1\,Gyr rather than 0.1-0.2\,Gyr (see also Girardi \& Bertelli 1998,
Marigo 1998). Consequently, the new integrated spectra evolve
from bluer to redder colours between 0.1 and 1\,Gyr rather than the opposite,
in better agreement with constraints from the clusters of the Magellanic
Clouds. One motivation for using the new models was to test whether
they would allow for a more recent onset of the intermediate age
star formation episode.

\begin{figure}[!h]
\centering
\includegraphics[angle=0,width=0.45\textwidth]{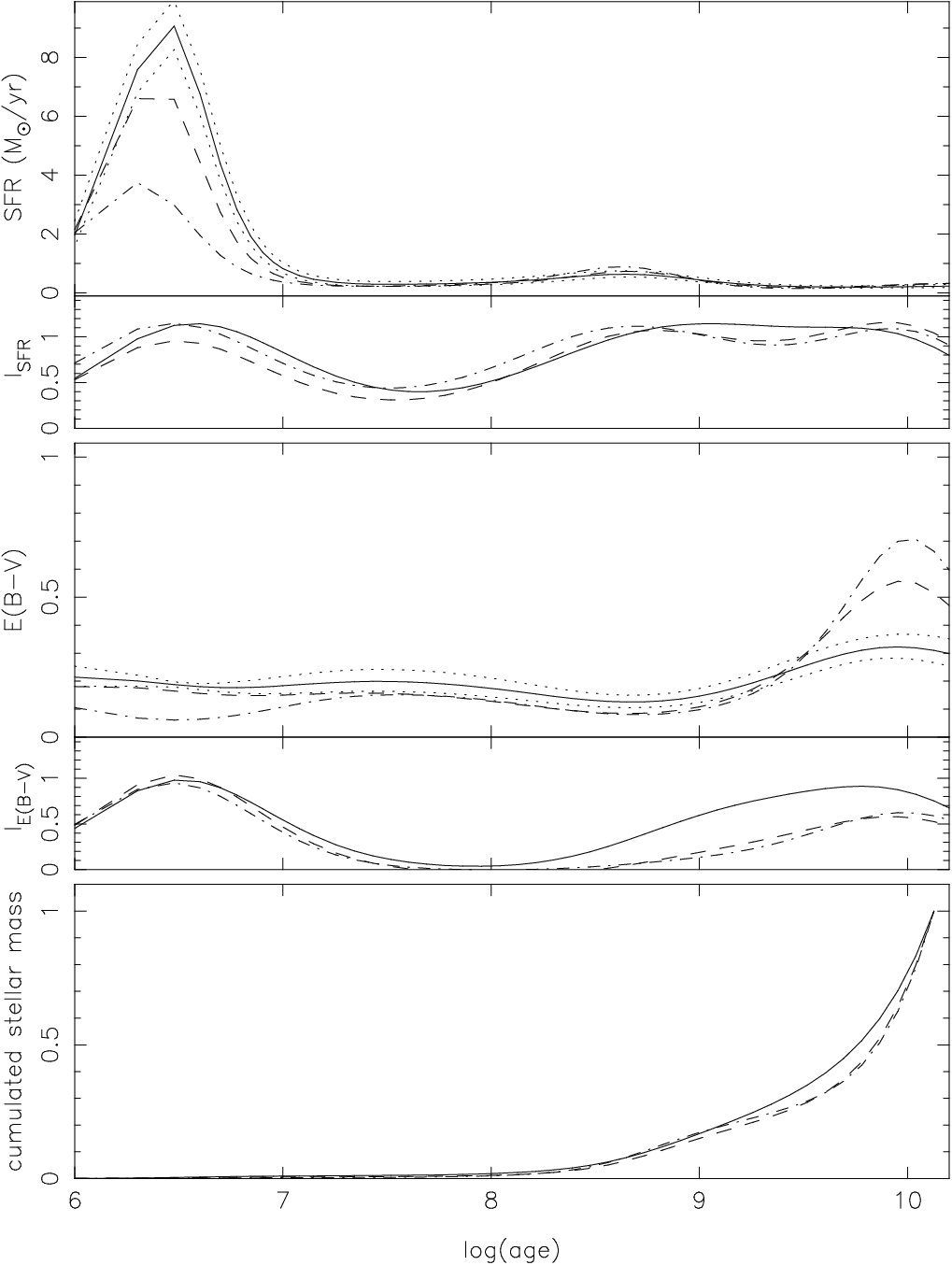}
\caption{Same as Fig.\,\ref{figext1}, but with the stellar populations
of Mouhcine \& Lan\c{c}on (2002).
Solid: Calzetti-type dust; dot-dashed: SMC dust screen;
dashed: dust clouds.
}
\label{figext2}
\end{figure}

Fig.\,\ref{figext2}  is the equivalent of Fig.\,\ref{figext1},
for the new basis of single population spectra.
The results are not significantly different. 
The intermediate
age star formation episode starts early again. In the nucleus
of NGC\,7714, because of the large contribution of the
old stars of the underlying spiral bulge, 
there is no need to invoke the reddest AGB-dominated 
populations that would differentiate between the two sets of 
basis spectra.

\subsection{Deriving the optical depth from the spectrum.}
\label{freelaw.sec}

The $\chi^2_{\nu}$ values of the previous models are of about 2, 
meaning that better models can be sought for. 
Moreover, the adjustment of the emission lines showed a specific problem: 
Br$\gamma$ is systematically underestimated by a factor of two. 
So, a possibility is that the extinction laws we have used do
not correspond to reality.
In this section, we derive a new extinction law, compatible 
with the data, emission lines included. 
More precisely, the model is written as follows:
\begin{equation}
F_{\lambda}= f_{ext}(\lambda) \int_{t_f}^{t_i} 
\psi_0 \, \exp(\alpha(t)) \, B_{\lambda}(t,Z_0)\,  {\rm d}t
\end{equation}
with:
\begin{equation}
f_{ext}(\lambda) = \exp(-\tilde{\tau}_{\lambda})
\end{equation}
Here, we consider the effective optical depth 
$\tilde{\tau}_{\lambda}$ as an unknown but assume
it is independent of age, and we adopt solar metallicity
for all stellar populations. We apply the same extinction
to the stellar continuum and to the nebular emission. The basis spectra used 
are those of Sect.\,\ref{extmod.sec}. Appendix \ref{opa} provides 
computational details and Table\,\ref{7714prior.tab} the priors.

\begin{figure}
\centering
\includegraphics[angle=0,width=0.45\textwidth]{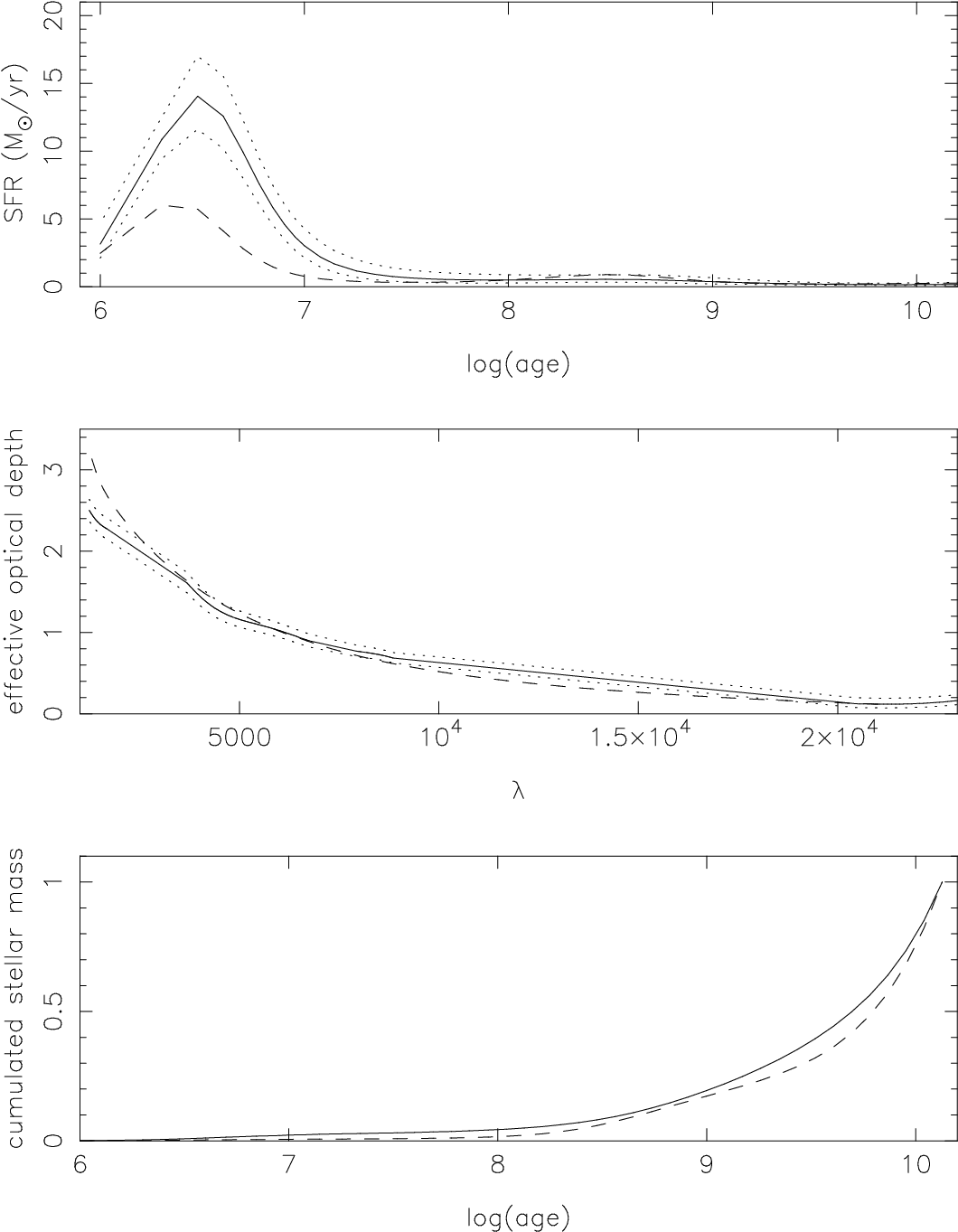}
\caption{Simultaneous determination of the star
formation rate (top panel) and of the optical depth (middle panel, full line)
for the central 300\,pc of NGC\,7714.
We have superimposed the Calzetti law, rescaled to $E(B-V)=0.3$ (dashed line).
The bottom panel shows the cumulative mass contribution of the stars.}
\label{figsalpfiocop1}
\end{figure}

\begin{figure}
\centering
%
\includegraphics[angle=0,width=0.45\textwidth]{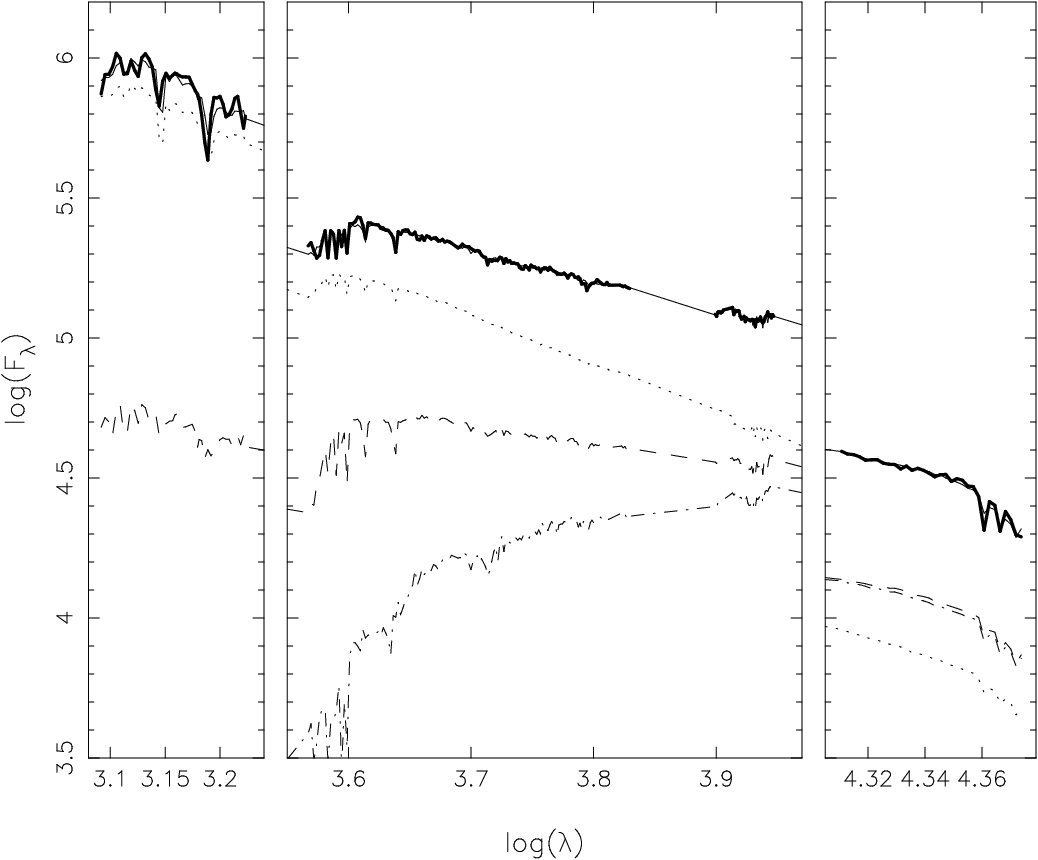}
\caption{Adjustment of the spectrum of the nucleus of NGC\,7714, 
with the SFR and attenuation obtained in Sect.\,\ref{freelaw.sec}. 
The line types are as in Fig.\,\ref{Calzfit.fig}.
}
\label{fitfreelaw}
\end{figure}

The SFR obtained here and shown in the top
panel of Fig.\,\ref{figsalpfiocop1}
provides a good fit to the data ($\chi^2_{\nu}$=1.4; Fig.\,\ref{fitfreelaw}).
The hydrogen lines are reproduced to within 10\,\%.
The mean index for the SFR is remarkably similar to
the curves plotted in Figs.\,\ref{figext1} or \ref{figext2}.
The mean index for $\tilde{\tau}_{\lambda}$ behaves as
described and explained in Sect.\,\ref{simul}.
The plotted attenuation law is derived to within an
additive constant, which corresponds to multiplicative 
factor on the SFR.  A minimum
level of the SFR is set by the constraint of positive extinction
at all wavelengths. If extinction were nil in the
near-IR, the SFR would be reduced by only $\exp(-0.1)$,
i.e. about 10\,\% as compared to the plotted curve.  
Much larger values of $\tilde{\tau}_{\lambda}$ 
and the SFR can be excluded when considering 
the far-IR emission of the galaxy, a constraint that we have
not incorporated in the inversion procedure. LGLG01 verified that,
with 4 times as many hot stars as directly derived from the UV flux 
with standard extinction laws, the nuclear $\sim 330$\,pc would 
already produce over a third of the total far-IR light of the galaxy.
This factor of 4 is already present in the 
SFR of Fig.\,\ref{figsalpfiocop1}.

Interstingly, the wavelength dependence of the derived attenuation is
close to the empirical law of Calzetti et al. (2000).
The differences, however, {\em are} significant: in the UV, the
estimated law is flatter (otherwise
the blue slope of the observed UV continuum could not
be reproduced with this amount of attenuation).
The present results mimic what imaging suggests
is the real situation in the central 300\,pc of
NGC\,7714 (LGLG01): a significant fraction of the
UV-emitting stars are heavily attenuated in the UV,
and contribute mainly to Br$\gamma$, while a smaller
fraction of these hot stars is practically unobscured,
explaining the small apparent reddening at UV wavelengths.

The SFR found here exceeds 1\,M$_{\odot}$\,yr$^{-1}$ only
in the last 20\,Myr. This timescale is consistent with the dynamical
models of Struck \& Smith (2002). Nevertheless, there is an
enhancement in the SFR about 7\,10$^8$\,yr ago (from about 0.1
to about 0.5\,M$_{\odot}$\,yr$^{-1}$).

Despite the appeal of this solution, we express
some caution. First, the fit of Fig.\,\ref{fitfreelaw}
shows the stars with ages below $10^7$\,yr as the 
strongest contributors around the Balmer jump. The fit
in this region is good because of a bump in the
obscuration law at the appropriate place. There is no physical
reason to expect such a bump. At higher spectral resolution,
the shape of Balmer line wings would be able to confirm the need for
a stronger contribution of intermediate age stars (LGLG01).
Second, we have made the strong assumption here
that a single empirical attenuation law applies to stellar populations
of all ages. The solutions of LGLG01,
based on the integrated spectrum {\em and} direct 
constraints from high resolution imaging,
assigned populations of different ages different apparent obscurations.
In a way, we have traded freedom in the wavelength dependence of
$\tilde{\tau}_{\lambda}$ against freedom in its time dependence.
Unfortunately, considering the quasi-infinite
possible combinations of dust properties and dust distributions in a
starburst, we doubt that any spatially integrated spectroscopic  data
will be able to fully break the degeneracies and make multiwavelength
images redundant.

\section{Conclusions}

In this paper we have presented an inversion method designed
to estimate the SFR, the AMR and the reddening properties 
of galaxies from their spectra. This method allows us not only to
derive best values for these functional parameters, but also to estimate 
the amount of information actually present in the data for each 
of the unknowns, as well as the posterior resolution in time,
depending on the S/N of the studied spectra. 

The main conclusions based on the inversion of simulated 
spectra are :

\begin{itemize}

  \item The SFR, the AMR and the reddening law are determined  
  with a resolution of about 0.3-0.5 in log(age) 
  for low spectral resolution UV\,+\,optical\,+\,near infrared spectra 
  when noise in the data is negligible (S/N=500).

  \item With S/N=25 and the same spectral coverage,
  the SFR is determined with a resolution of 
  $\sim$0.6 in log(age) and the information on the AMR is
  rather poor. Constraints on the reddening of the stellar 
  continuum emission are tightest for the young stellar populations.
  The exact amount of recoverable information will depend
  on the actual star formation history of the object of study. 
  Similar success has been 
  reported by Cid Fernandes et al. (2001) in the framework of 
  empirical population synthesis.

  \item Alternatively, it is possible to constrain the SFR and the 
  wavelength dependence of the effective optical depth 
  simultaneously with a S/N=25, if it is assumed that this optical
  depth is independent of age (i.e. applies to all stellar populations)
  and that the stellar metallicity is known. In that
  case, the resolution on the SFR is only about 1 in log(age). 

\end{itemize} 

The inversion method applied to the spectrum of 
NGC 7714 confirms that the reddening 
properties of this galaxy cannot be modeled with simple extinction 
models. In order to obtain satisfactory fits to both the stellar
energy distribution and the nebular emission lines, one has to 
allow the wavelength dependence of attenuation to deviate from
``standard" laws (this work) or to allow for variations in the attenuation
between different coexisting stellar populations (LGLG01).

The main results obtained for NGC 7714 can be compared to 
the previous results from non-automated studies (LGLG01):

\begin{itemize}

\item The SFR found is indeed multimodal, with a young star formation
peak a few Myr ago and a more extended episode 
that formed stars at a typical rate of 0.5-1\,M$_{\odot}$\,yr$^{-1}$,
thus providing an important intermediate age population.

\item No solution was found in which the episode of enhanced star 
formation that produced the bulk of the intermediate age stars started less 
than $\sim$\,300\,Myr ago. The discrepancy with the dynamical results
of Smith \& Wallin (1992), who estimate that $\sim 100$\,Myr
elapsed since closest encounter with NGC\,7715, thus persists.
New simulations tend to confirm the dynamical timescales
(Struck \& Smith 2002), although a wider exploration of initial
conditions may still allow it to be stretched a little. The importance
of the event in terms of the stellar mass produced, especially
when compared to the most recent starburst, tends to favour
the occurence of an earlier perturbation, possibly unrelated with
NGC\,7715.

\item By solving simultaneously for the SFR and the effective optical depth
one strongly increases the estimated amplitude of the youngest peak in the SFR,
while producing an attenuation law that is flattened in the UV part 
of the spectrum. The global shape of the attenuation is similar to 
the average law derived by Calzetti et al. (2000), but the differences 
in the details are significant when the star formation rates are studied.
The Br$\gamma$ line emission is well matched, while it was
underestimated by a factor of 2 with ``classical" attenuation laws.
A drawback of this solution is that it produces a suspicious bump
in the optical obscuration law in order to match the shape of 
the Balmer jump. 

\item The solution just described mimics the effects that 
LGLG01 have shown to be due to a very inhomogeneous
distribution of the dust within the small region observed,
with one particular line of sight of very low extinction towards a
cluster of hot stars. Even with UV, optical and near-IR spectroscopy
combined, solving for the effective optical depths requires strong
assumptions. Here we have traded freedom in the wavelength dependence
of optical depths against the freedom to assign populations of
different ages differing amounts of extinction. Nature might be 
even more complex, and require freedom in both when individual
galaxies are studied. 

\end{itemize}

The precious information of the spatial
distribution of stars and dust within the region observed spectroscopically
is not always accessible. The detailed study of objects for
which high resolution imaging is possible remains essential, and
gives us insight into the fundamental uncertainties involved when
only integrated light is available. Improvements will come
naturally from a wider spectral coverage (if possible with overlaps
between spectral ranges observed independently), and from the
use of models with higher spectral resolution.

\begin{acknowledgements}
We thank an anonymous referee for a careful reading
and for requests that significantly improved the paper.
\end{acknowledgements}

\appendix

\section{Dust clouds and Poisson law.}
\label{dust_clouds.sec}
When clouds are distributed homogeneously in space we can describe 
correctly the extinction as due to a Poisson distribution of clouds.
The quantity $\overline{n}(t)$ represents the mean number of clouds 
encountered on the line of sight. 

We assume here that the clouds are of a typical size 
independent of age, so that the column density per cloud $E(B-V)_c$ is constant. 
In other words, we assume that the variation of the 
extinction in time depends only on the density of clouds.

The probability to find $i$ clouds in a given direction will follow 
a Poisson law :
\begin{equation}
P(i)=e^{-\overline{n}(t)} \frac{\overline{n}(t)^i}{i!}
\end{equation}
The proportion of the stellar population that is reddened by $i$ clouds is  
$P(i)$. The absorption by $i$ clouds is $e^{-i k_{\lambda} E(B-V)_c}$. 
Then, the resulting flux is (Natta and Panagia 1984) :
\begin{equation}
F_{\lambda}=\sum_{i=0}^{\infty}
\int_0^{t_i} \psi(t)\, B_{\lambda}(t,Z(t))\, 
P(i) \,e^{-i k_{\lambda} E(B-V)_c} \, {\rm d}t
\end{equation}
That yields :
\begin{eqnarray}
F_{\lambda} \ = \int_0^{t_i} \psi(t) B_{\lambda}(t,Z(t)) \,
 e^{-\overline{n}(t)} \sum_{i=0}^{\infty}
  \frac{\overline{n}(t)^i}{i!} e^{-i k_{\lambda} E(B-V)_c} \, {\rm d}t
 	\nonumber \\
&
\end{eqnarray}
with :
\begin{equation}
\sum_{i=0}^{\infty}\frac{\overline{n}(t)^i}{i!} e^{-i k_{\lambda} E(B-V)_c}=
e^{\overline{n}(t) e^{-\tau_{\lambda ,c}}}
\end{equation}

\section{Estimation of the parameters.}

\subsection{Change of variable: $u=\log(t)$.}
\label{varchange}
Massive stars evolve more rapidly than small mass stars, and
a given absolute difference in initial masses has more effect
on stellar evolution in the high mass regime than in the low mass one.
As a consequence, the time interval over which spectrophotometric properties of
isochrone stellar populations vary significantly increases
quasi-linearly with age. 
The resolution in time expected for the SFR is thus almost
constant in $\log(t)$ (see references in 
Hubeny et al. 1999, and Lan\c{c}on 2000).
The following change of variable is natural:
\begin{equation}
u=\log(t)
\end{equation}
The fundamental equation (Equation\,\ref{gen}) should be written as follows:
\begin{equation}
F_{\lambda}=\int_{\log(t_f)}^{\log(t_i)} \psi(u) \, B_{\lambda}(u,Z(u)) \,
f_{ext}(\lambda,u) \, h(u) \, {\rm d}u
\end{equation}
with $h(u)=10^u {\rm ln}(10)$.

So, the prior variance-covariance operators defined below ($C_{\alpha}$, 
$C_{Z}$, ...etc.) are expressed with the new variable $u$. For instance, 
$C_{\alpha}$ is written as :
\begin{equation}
C_{\alpha}(u,u')=\sigma_{\alpha}(u) \sigma_{\alpha}(u') 
\exp\left(- \frac{(u-u')^2}{\xi_{\alpha}^2} \right)
\end{equation}
so $\xi_{\alpha}$ is expressed in units of log($t$).

\subsection{Determination of $\psi$, $Z$, $\overline{n}$ and $E(B-V)$}
\label{pardet} 

In this section, we clarify the main steps of the inversion
procedure, in the case one wishes to derive the time 
dependence of $\psi$, $Z$ and $\overline{n}$, and the
value of $E(B-V)$ (the extinction per cloud on the line
of sight).

Since the star formation rate is always positive, changing over from 
$\psi(u)$ to $\alpha(u)$ in Equation\,\ref{gen} yields:
\begin{equation}
\alpha(u) := \ln(\psi(u)/\psi_0)
\end{equation}
where $\psi_0$ is a constant. We shall assume that the uncertainties 
in the function $\alpha(u)$ follow a Gaussian law. This implies that the 
errors in $\psi(u)$ follow a log-normal distribution.
Similarly, the equations can be rewritten with logarithmic
variables for all unknowns that must be forced to positive values.

The flux at the wavelength $\lambda_i$ is modeled as follows :
\begin{eqnarray}
F_{\lambda_i}& = & \int_{\log(t_f)}^{\log(t_i)} \psi_0 \, B_{\lambda_i}(u,Z(u)) 
		 \nonumber \\
& & \  \exp\left\{\alpha(u)+\overline{n}(u)  
 \left(e^{-k_{\lambda_i} E(B-V)}-1\right)\right\} \, h(u)\, {\rm d}u 
			\nonumber \\
& = & g_{\lambda_i}(\alpha(u),Z(u),\overline{n}(u),E(B-V))
\end{eqnarray}
This formulation requires use of the inverse method in the non-linear
case (Equation\,\ref{min2}). We drop the $[k]$ index in order to simplify
the notation.\\
The vector $M$ of unknown parameters is :
\begin{equation}
      M  = \left( \begin{array}{c}
                     \alpha(u)        \\
                     Z(u)             \\
                     \overline{n}(u)  \\
                     E(B-V)                \\
                  \end{array}
           \right)
\end{equation}
The prior variance-covariance matrix shall be written :
\begin{equation}
    C_0=\left(\begin{array}{cccc}
                C_{\alpha}   & 0  & 0 & 0        \\
                0            & C_Z & 0 & 0 \\
                0  & 0   & C_{\overline{n}} & 0  \\
                0  & 0   & 0 & \sigma^2_{E(B-V)}  \\
              \end{array}
        \right)
\label{Co_equation}
\end{equation}
Here $C_{\alpha}$ (for instance) contains diagonal terms of
the form $C_{\alpha}(u,u)=\sigma_{\alpha}^2(u)$ and non-diagonal 
terms of the form $C_{\alpha}(u,u')$\ (see Equation\,\ref{corsfr}). 

The matrix of partial derivatives is (with $s$ the number of 
wavelengths sampled):
\begin{equation}
     G=\left( \begin{array}{cccc}
                   \frac{\partial g_{\lambda_1}}{\partial \alpha}
                        &\frac{\partial g_{\lambda_1}}{\partial Z}
                        &\frac{\partial g_{\lambda_1}}{\partial \overline{n}}
                        &\frac{\partial g_{\lambda_1}}{\partial E(B-V)} \\
                   \frac{\partial g_{\lambda_2}}{\partial \alpha}
                        &\frac{\partial g_{\lambda_2}}{\partial Z}
                        &\frac{\partial g_{\lambda_2}}{\partial \overline{n}}
                        &\frac{\partial g_{\lambda_2}}{\partial E(B-V)} \\
                   \vdots      & \vdots & \vdots & \vdots \\
                   \frac{\partial g_{\lambda_s}}{\partial \alpha}
                        &\frac{\partial g_{\lambda_s}}{\partial Z}
                        &\frac{\partial g_{\lambda_s}}{\partial \overline{n}}
                        &\frac{\partial g_{\lambda_s}}{\partial E(B-V)} \\
                \end{array}
          \right)
\end{equation}

$E(B-V)$ being a discrete parameter, the corresponding partial derivative is not
a functional operator.
$\partial g_{\lambda}/ \partial E(B-V)$ is written:
\begin{eqnarray}
\lefteqn{ \frac{\partial g_{\lambda}}{\partial E(B-V)}  = 
		} \nonumber \\
& & -\int_{\log(t_f)}^{\log(t_i)} \psi_0\,B_{\lambda}(u,Z(u)) 
 \,\, \overline{n}(u) k_{\lambda} e^{-k_{\lambda} E(B-V)}
		\nonumber \\
& & \quad \exp\left\{\alpha(u)+\overline{n}(u) 
	\left(e^{-k_{\lambda} E(B-V)}-1\right)\right\}\, h(u) \,{\rm d}u
\end{eqnarray}

Conversely, $\alpha$, $Z$ and $\overline{n}$ are functions of 
the time variable. The corresponding partial derivatives are
functional operators.
$\frac{\partial g_{\lambda}}{\partial \alpha}$, 
$\frac{\partial g_{\lambda}}{\partial Z}$ and  
$\frac{\partial g_{\lambda}}{\partial \overline{n}}$
have respectively the following kernels : 
\begin{eqnarray}
\lefteqn{ K_1(u)  =  \psi_0 \, B_{\lambda}(u,Z(u)) } \nonumber \\
   & \qquad \exp\left\{\alpha(u)+\overline{n}(u)
		\left(e^{-k_{\lambda} E(B-V)}-1\right)\right\} h(u) 
\end{eqnarray}
\begin{eqnarray}
\lefteqn{ K_2(u)  =  \psi_0 \, \frac{\partial B_{\lambda}(u,Z(u))}{\partial Z}
				}	\nonumber \\ 
   & \qquad \exp\left\{\alpha(u)+\overline{n}(u)
		\left(e^{-k_{\lambda} E(B-V)}-1\right)\right\} h(u) 
\end{eqnarray}
\begin{eqnarray}
\lefteqn{ K_3(u)  =  \left(e^{-k_{\lambda} E(B-V)}-1\right)
		\, \psi_0 \, B_{\lambda}(u,Z(u)) } \nonumber \\
   & \qquad \exp\left\{\alpha(u)+\overline{n}(u)
		\left(e^{-k_{\lambda} E(B-V)}-1\right)\right\} h(u) 
\end{eqnarray}

These kernels $K_i(u)$  act on a function $f(u)$ as follows:
\begin{equation}
\label{scalproduct}
<K_i,f>=\int_{\log(t_f)}^{\log(t_i)} f(u) K_i(u) {\rm d}u
\end{equation}
where $<,>$ is the scalar product in $L_2$ Hilbert space.
\smallskip

The $i$-th component of the vector $V_i=D + G \cdot
(M -M_0)-g(M)$ (in Equation\,\ref{min2}) is:
{\setlength\arraycolsep{2pt}  
\begin{eqnarray}
V_i & = & F_{\lambda_i}
 + \frac{\partial g_{\lambda_i}}{\partial \alpha} (\alpha(u)-\alpha_0) 
				\nonumber \\
& & + \frac{\partial g_{\lambda_i}}{\partial Z} (Z(u)-Z_0)  
  + \frac{\partial g_{\lambda_i}}{\partial \overline{n}}(\overline{n}(u)-
    \overline{n}_0)   \nonumber \\
& & +\frac{\partial g_{\lambda_i}} {\partial E(B-V)} \big(E(B-V)-E(B-V)_0\big)
			\nonumber \\
& & - \, g_{\lambda_i}\big(\alpha(u),Z(u),\overline{n}(u),E(B-V)\big)
\end{eqnarray} }
where $\alpha_0$,$Z_0$, $\overline{n}_0$ and $E(B-V)_0$ 
are respectively the priors 
of $\alpha(u)$, $Z(u)$, $\overline{n}(u)$ and $E(B-V)$.
The $(i,j)$-th component of $(G C_0 G^{*} + C_d)$ is:
\begin{eqnarray}
\label{mats}
S_{i,j} &= &
\frac{\partial g_{\lambda_i}}{\partial \alpha} \,C_{\alpha}
\frac{\partial g_{\lambda_j}}{\partial \alpha}+ 
\frac{\partial g_{\lambda_i}}{\partial Z} \,C_Z 
\frac{\partial g_{\lambda_j}}{\partial Z}+
\frac{\partial g_{\lambda_i}}{\partial \overline{n}}\, C_{\overline{n}}\,
\frac{\partial g_{\lambda_j}}{\partial \overline{n}}
		\nonumber \\
& & + \frac{\partial g_{\lambda_i}}{\partial E(B-V)} \sigma^2_{E(B-V)} 
\frac{\partial g_{\lambda_j}}{\partial E(B-V)} 
+\delta_{i,j}\,\sigma_i\,\sigma_j \nonumber \\
& &
\end{eqnarray}
where $\delta_{i,j}$ is the Kronecker symbol and $\sigma_i$ the 
root mean square of the noise in the data $F_{\lambda_i}$ (here we assume
uncorrelated noise in the data).

Defining the vector : 
\begin{equation}
\label{vecw}
W=S^{-1}V,
\end{equation}
the estimation number $(k+1)$ for the 
parameters is computed from the previously estimated values by :
\begin{eqnarray}
\alpha_{[k+1]}(u)=\alpha_0+\sum_i W_i \int_{\log(t_f)}^{\log(t_i)} 
C_{\alpha}(u,u')K_1(u') {\rm d} u' \nonumber \\
& \\ 
Z_{[k+1]}(u)=Z_0 + \sum_i W_i \int_{\log(t_f)}^{\log(t_i)} 
C_Z(u,u')K_2(u') {\rm d} u' \nonumber \\
& \\
\overline{n}_{[k+1]}(u)=\overline{n}_0+ \sum_i W_i \int_{\log(t_f)}^{\log(t_i)} 
C_{\overline{n}}(u,u') K_3(u')  {\rm d} u' \nonumber \\
& 
\end{eqnarray}
and  
\begin{eqnarray}
E(B-V)_{[k+1]}& =& E(B-V)_0 \nonumber \\
	& & +\, \sigma_{E(B-V)}^2 \sum_i W_i 
	\frac{\partial g_{\lambda_i}}{\partial E(B-V)} 
\end{eqnarray}

\subsection{Determination of $\psi$ and $\tilde{\tau}_{\lambda}$.}
\label{opa}
In this section, the model and the observations are linked by 
the relation :
\begin{eqnarray}
F_{\lambda_i}& =&  \exp\left(-\tilde{\tau}_{\lambda_i} \right)
 \int_{\log(t_f)}^{\log(t_i)} \psi_0 B_{\lambda_i}(u,Z_0) 
 e^{\alpha(u)}\, h(u) {\rm d}u  \\
& = & g_{\lambda_i}(\alpha(u),\tau_{\lambda}) \nonumber
\end{eqnarray}
where the metallicity is fixed. 
The unknown parameters are $\alpha(u)$ and $\tilde{\tau}_{\lambda}$ with 
the respective variance-covariance operators
$C_{\alpha}$ and $C_{\tilde{\tau}}$.\\
\begin{equation}
      M  = \left( \begin{array}{c}
                     \alpha(u)        \\
                     \tilde{\tau}_{\lambda}   \\
                  \end{array}
           \right)
\end{equation}
\begin{equation}
    C_0=\left(\begin{array}{cc}
                C_{\alpha}   & 0        \\
                0            & C_{\tilde{\tau}} \\
              \end{array}
        \right)
\end{equation}
with :
\begin{equation}
C_{\tilde{\tau}}(\lambda,\lambda')=
\sigma_{\tau}(\lambda) \sigma_{\tau}(\lambda') 
\exp\left( -\frac{(\lambda-\lambda')^2}{\xi_{\tilde{\tau}}^2}\right)
\end{equation}
$C_{\alpha}$ is given by Equation\,\ref{corsfr}.
To force positive values of $\tilde{\tau}_{\lambda}$, the 
equations are rewritten with ln($\tilde{\tau}_{\lambda}$) as
the unknown.

The matrix of partial derivatives of Equation\,\ref{min2} is :
\begin{equation}
     G=\left( \begin{array}{cc}
                   \frac{\partial g_{\lambda_1}}{\partial \alpha}
                        &\frac{\partial g_{\lambda_1}}{\partial \tilde{\tau}}\\
                   \frac{\partial g_{\lambda_2}}{\partial \alpha}
                        &\frac{\partial g_{\lambda_2}}{\partial \tilde{\tau}}\\
                        \vdots      & \vdots \\
                   \frac{\partial g_{\lambda_s}}{\partial \alpha}
                        &\frac{\partial g_{\lambda_s}}{\partial \tilde{\tau}}\\
                \end{array}
          \right)
\end{equation}
$\frac{\partial g_{\lambda}}{\partial \alpha}$, 
$\frac{\partial g_{\lambda}}{\partial \tilde{\tau}}$  
respectively have the following kernels :
\begin{eqnarray}
K_1(u)&=&\psi_0 \exp\left(\alpha(u)\!-\!\tilde{\tau}_{\lambda}\right)
B_{\lambda}(u,Z_0) h(u) \\
K_2(\lambda)&=&\psi_0 \exp\left(\alpha(u)\!-\!\tilde{\tau}_{\lambda}\right)
B_{\lambda}(u,Z_0) \delta(\lambda\!-\!\lambda_i) h(u)
\end{eqnarray} 
where $\delta$ is the Dirac function. These kernels act on 
a function as shown in Equation\,\ref{scalproduct}.

The $i$-th component of the vector $V_i=D + G \cdot
(M -M_0)-g(M)$ is :
\begin{eqnarray}
V_i& =& F_{\lambda_i}
 +\frac{\partial g_{\lambda_i}}{\partial \alpha} (\alpha(u)-\alpha_0)\
 + \frac{\partial g_{\lambda_1}}{\partial \tilde{\tau}}
	(\tilde{\tau}_{\lambda}-\tilde{\tau}_{\lambda,0})   
	\nonumber \\
& & - g_{\lambda_i}(\alpha(u),\tilde{\tau}_{\lambda})
\end{eqnarray}
where $\alpha_0$ and $\tilde{\tau}_{\lambda,0}$ are respectively the priors 
of $\alpha(u)$ and $\tilde{\tau}_{\lambda}$.\\
The matrix $S$ (Equation\,\ref{mats}) is computed as follows:
\begin{equation}
S_{i,j} = 
\frac{\partial g_{\lambda_i}}{\partial \alpha} C_{\alpha}
\frac{\partial g_{\lambda_j}}{\partial \alpha}+
\frac{\partial g_{\lambda_i}}{\partial \tilde{\tau}_{\lambda}} C_{\tilde{\tau}}
\frac{\partial g_{\lambda_j}}{\partial \tilde{\tau}_{\lambda}}
+\delta_{i,j}\,\sigma_i\,\sigma_j
\end{equation}
where $W$ is a vector defined by Equation\,\ref{vecw}.
The estimation of the parameters is given by :
\begin{equation}
\alpha_{[k+1]}(u)=\alpha_0+\!\sum_i W_i \int_{\log(t_f)}^{\log(t_i)}
 \!\!C_{\alpha}(u,u')K_1(u') {\rm d} u'
\end{equation}
\begin{equation}
\tilde{\tau}_{\lambda[k+1]}=\tilde{\tau}_{\lambda,0} + \sum_i W_i \int_0^{\infty}
C_{\tilde{\tau}}(\lambda,\lambda')K_2(\lambda') {\rm d} \lambda'
\end{equation}

\end{document}